\documentclass[aps,pre,twocolumn,showpacs,superscriptaddress,groupaddress]{revtex4-1}
\usepackage[breaklinks=true]{hyperref}
\usepackage{amsmath}
\usepackage{amsfonts}
\usepackage{amssymb}
\usepackage{epsfig}
\usepackage{graphicx}
\usepackage[usenames,dvipsnames]{color}
\usepackage{bm}
\usepackage{float}
\usepackage{tabularx}
\usepackage[utf8]{inputenc}

\definecolor{snapyellow}{RGB}{204,188,41}
\definecolor{snapred}{RGB}{204,41,41}
\definecolor{apsblue}{rgb}{0.18,0.19,0.57}

\hypersetup{
 colorlinks,
 citecolor=apsblue,
 urlcolor=apsblue,
 linkcolor=apsblue,
 pdfpagemode={UseNone},
 pdfauthor={},
 pdftitle={},
}

\newcommand{\beq}{\begin{equation}}
\newcommand{\eeq}{\end{equation}}
\newcommand{\nbar}{\overline{n}}

\begin{document}
\title{Assessing the structural heterogeneity of supercooled liquids through community inference}

\author{Joris Paret}
\affiliation{Laboratoire Charles Coulomb (L2C), Universit\'e de Montpellier, CNRS, Montpellier, France}
\author{Robert L. Jack}
\affiliation{Department of Chemistry, University of Cambridge, Lensfield Road, Cambridge CB2 1EW, United Kingdom}
\affiliation{Department of Applied Mathematics and Theoretical Physics, University of Cambridge, Wilberforce Road, Cambridge CB3 0WA, United Kingdom}
\author{Daniele Coslovich}
\thanks{This article may be downloaded for personal use only. Any other use requires prior permission of the author and AIP Publishing. This article appeared in The Journal of Chemical Physics and may be found at \url{https://doi.org/10.1063/5.0004732}.}
\affiliation{Laboratoire Charles Coulomb (L2C), Universit\'e de Montpellier, CNRS, Montpellier, France}

\begin{abstract}
We present an information-theoretic approach inspired by distributional clustering to assess the structural heterogeneity of particulate systems. Our method identifies communities of particles that share a similar local structure by harvesting the information hidden in the spatial variation of two- or three-body static correlations.  This corresponds to an unsupervised machine learning approach that infers communities solely from the particle positions and their species.  We apply this method to three models of supercooled liquids and find that it detects subtle forms of local order, as demonstrated by a comparison with the statistics of Voronoi cells. Finally, we analyze the time-dependent correlation between structural communities and particle mobility and show that our method captures relevant information about glassy dynamics.
\end{abstract}

\maketitle

\section{Introduction}

The viscosity and structural relaxation times of supercooled liquids increase by several orders of magnitude on approaching the glass transition.
Remarkably, such a drastic slowing down of the dynamics occurs without a marked change of the local structure: conventional correlation functions such as the static structure factor look qualitatively similar in the liquid and in the glass~\cite{Berthier_Biroli_2011}.
One possible explanation for this disconnect is that the structural features relevant to the glassy slowdown are hard to detect in two-body correlations.
This is illustrated, for instance, by the appearance of particle arrangements with icosahedral symmetry, whose first evidence in simple glassy mixtures dates back to the pioneering numerical studies of J\'onsson and Andersen~\cite{Jonsson_Andersen_1988}.
Later on, short range icosahedral order was reported in metallic alloys~\cite{Luo_Sheng_Alamgir_Bai_He_Ma_2004} and colloidal suspensions~\cite{Tanaka_Kawasaki_Shintani_Watanabe_2010}.
Over the last years, computer simulations and experiments revealed the presence of more general locally stable motifs, known as ``locally favored structures'' (LFS), which
can be detected using the Voronoi tessellation~\cite{Sheng_Luo_Alamgir_Bai_Ma_2006,coslovichUnderstandingFragilitySupercooled2007a}, topological cluster classification~\cite{malinsIdentificationStructureCondensed2013,malinsIdentificationLonglivedClusters2013,malinsLifetimesLengthscalesStructural2014}, bond orientational order analysis~\cite{leocmachRolesIcosahedralCrystallike2012,leocmachImportanceManybodyCorrelations2013} and alternative approaches~\cite{fangAtomisticClusterAlignment2010,keysCharacterizingComplexParticle2011}.
The emergence of well-defined locally favored structures suggests a reduction of structural diversity compared to the normal liquid~\cite{weiAssessingUtilityStructure2019} and hints to enhanced spatial variations of the preferred local order~\cite{royallRoleLocalStructure2015}.

To assess whether a given measure of local order has a definite link to the dynamics of the particles, Harrowell and coworkers introduced the isoconfigurational ensemble~\cite{widmer-cooperHowReproducibleAre2004}, which prescribes a statistical average over an ensemble of trajectories initiating from the configuration of interest.
This approach effectively filters out dynamic fluctuations irreproducible from a given particle configuration.
The correlation between the spatial fluctuations of dynamics in the isoconfigurational ensemble and local structural descriptors appears, however, system-dependent~\cite{hockyCorrelationLocalOrder2014,coslovichStructureInactiveStates2016}.
Recent studies have shown that order parameters quantifying local packing efficiency~\cite{tongRevealingHiddenStructural2018,Marin-Aguilar_Wensink_Foffi_Smallenburg_2019,Tong_Tanaka_2019} are highly predictive of the dynamics in hard (or nearly hard) spheres, but these results do not carry over universally to other models.
The spatial distribution of soft modes~\cite{widmer-cooperIrreversibleReorganizationSupercooled2008,widmer-cooperPredictingLongTimeDynamic2006} correlates well with the local dynamics, at least on time scales shorter than the structural relaxation time, but normal modes obviously contain richer information than the bare structure, since they account for local variations of the energy function.
Machine learning techniques have also been used to identify structural defects related to localized excitations in supercooled liquids, as well as plastic events in glasses~\cite{cubukIdentifyingStructuralFlow2015,schoenholzStructuralApproachRelaxation2016,cubukStructuralPropertiesDefects2016,cubukStructurepropertyRelationshipsUniversal2017,maHeterogeneousActivationLocal2019}. While promising, supervised approaches still need input dynamic data to identify relevant structural features.

In this work, we describe a method for identifying local order, based on statistical inference, without prior knowledge of dominant packing motifs or LFS.
Instead of characterizing order with complex geometrical fingerprints, such as in cluster identification or bond-order analysis, the method works with much simpler quantities -- interparticle distances and bond angles.  The key idea is that particles in locally ordered environments are characterized by non-typical distributions of their neighbors.
We use this fact to group particles into \emph{structural communities} sharing a similar local structure, which in turn differs markedly from the one of the other communities.
In the context of machine learning and statistical inference, this is an example of a clustering analysis method~\cite{HastieBook}.

Our specific implementation builds on simple information-theoretic concepts~\cite{coverElementsInformationTheory2006}, which have recently found fruitful application in studies of supercooled liquids~\cite{dunleavyUsingMutualInformation2012,jackInformationTheoreticMeasurementsCoupling2014,dunleavyMutualInformationReveals2015}, and on the spatial fluctuations of two- and three-body static correlations.
Clustering analysis methods based on similar information-theoretic ideas have also been applied in other contexts~\cite{Dhillon2003,Faiv2010}.
Our approach differs from network-theoretic community detection~\cite{newmanNetworksIntroduction2010}, which has been used to investigate the structure of models of supercooled liquids~\cite{ronhovdeMultiresolutionCommunityDetection2009,ronhovdeLocalResolutionlimitfreePotts2010,ronhovdeDetectingHiddenSpatial2011,ronhovdeLocalMultiresolutionOrder2015} and to determine force networks in granular materials~\cite{bassettExtractionForcechainNetwork2015}.
In that context, network-theoretic community detection identifies groups of particles (nodes) that are tightly connected to one another, with couplings that weight the proximity of particles using energy terms from the underlying particle model or from the radial distribution function (RDF).
By contrast, the structural communities discussed here do not imply a priori a notion of physical neighborhood in real space~\footnote{For this reason we also refrain from using the term ``cluster'', which carries an even more obvious notion of spatial proximity between particles.}.
We will show that these communities still convey relevant information on the spatial fluctuations of the dynamics, especially over length scales of the order of the interparticle distance.

The paper is organized as follows.
In Sec.~\ref{sec:theory} we introduce the basic theoretical concepts and methods to identify structural communities using the mutual information between communities and local structural descriptors.
In Sec.~\ref{sec:num-methods}, we describe the models and the numerical implementation of the methods. Section~\ref{sec:results-geom} identifies the main features of the structural communities.
In Sec.~\ref{sec:corr-struct-dyn} we assess the correlation between the dynamics in the isoconfigurational ensemble and the spatial fluctuations of the communities.
In Sec.~\ref{sec:discussion} we provide an overall assessment of the method and we conclude in Sec.~\ref{sec:conclusions} by suggesting possible extensions and improvements.

\section{Information-theoretic inference of communities}
\label{sec:theory}

\subsection{Overview and motivation}

We present an algorithm that characterizes structural heterogeneity in a particulate system.  We identify structural communities such that particles in the same community have similar local structure.  For example, imagine a structurally heterogeneous liquid in which some particles have highly-ordered local environments, while others are more disordered.
These particles could be separated into two communities, according to their local order.
The aim of our method is to identify such communities using statistical inference, with minimal prior assumptions on the nature of the local order.
Our method is unsupervised and uses only structural information -- this is distinct from other inference or machine learning approaches that learn about dynamically-active regions or soft spots using training data sets~\cite{cubukIdentifyingStructuralFlow2015,schoenholzStructuralApproachRelaxation2016,cubukStructuralPropertiesDefects2016,cubukStructurepropertyRelationshipsUniversal2017,maHeterogeneousActivationLocal2019}.

The communities are labeled by $k$ (with $0\leq k \leq {\cal K}-1$) and $s$ is a property characterizing a particle and its neighborhood.
For example, $s$ might be the interparticle distance, \textit{i.e.}, the distance between a particle and one of its neighbors.
We write the joint distribution of $k$ and $s$ as $p(k,s)$ and we note that $p(k,s) = f_k p_k(s)$, where $f_k$ is the fraction of particles in community $k$ and $p_k(s)$ is the distribution of $s$ for particles in community $k$.
Also let $p(s) = \sum_k p(k,s)$ be the marginal distribution of $s$. 
In the example where $s$ is an interparticle distance, $p_k(s)$ is the distribution of distances for community $k$ and $p(s)$ is the full distribution of distances.

In order to have meaningful communities, the distributions $p_0,p_1,\dots,p_{{\cal K}-1}$ should all differ significantly from each other.  
To quantify this, we consider
the mutual information (MI)~\cite{coverElementsInformationTheory2006} between $k$ and $s$:
\beq
\label{eq:MIka1}
I(k;s) = \sum_{k=0}^{\mathcal{K}-1} \int p(k,s) \log \left( \frac{ p(k,s) }{ f_k p(s) } \right) \mathrm{d}s ,
\eeq
This MI is the amount of information about a particle's value of $s$ that is provided by a measurement of its community label $k$.
The choice of base for the logarithm fixes the units of MI -- in numerical work we use logarithms in base 2 so that $I$ is expressed in bits, but the general theory is independent of this basis.
The MI is symmetric, so it is also equal to the amount of information about $k$ that is provided by a measurement of $s$.
This MI is large if $k$ and $s$ are strongly correlated with each other, in which case the communities differ significantly.  
To make this apparent we rewrite Eq.~\eqref{eq:MIka1} as
\beq
\label{eq:MIka}
I(k;s) = \sum_{k=0}^{\mathcal{K}-1} \int f_k p_k(s) \log \left( \frac{ p_k(s) }{ p(s) } \right) \mathrm{d}s \;.
\eeq

The essence of our community inference method is thus to maximize the mutual information $I(k;s)$, for a given structural measure $s$.
By doing so, particles sharing a similar local structure will tend to be clustered into the same community.
As we will see in the next section, the method that we describe can infer communities from different kinds of structural measures $s$, which makes it general and versatile.
We also describe an extension of the method that accounts for the fluctuations of additional static fields, such as local density and composition, between communities.
Through this extension we shall make contact with related approaches based on spatially resolved two-body entropy~\cite{leocmachImportanceManybodyCorrelations2013,Piaggi_Parrinello_2017}.

\subsection{Community inference method}

In this section we present the community inference method in its simplest form. We start by considering a simple fluid where all particles are treated as identical.
The extension to the practically-relevant case of multi-component mixtures is discussed in Sec.~\ref{sec:liquid-mix}.

\begin{figure*}[!t]
	\includegraphics[scale=1]{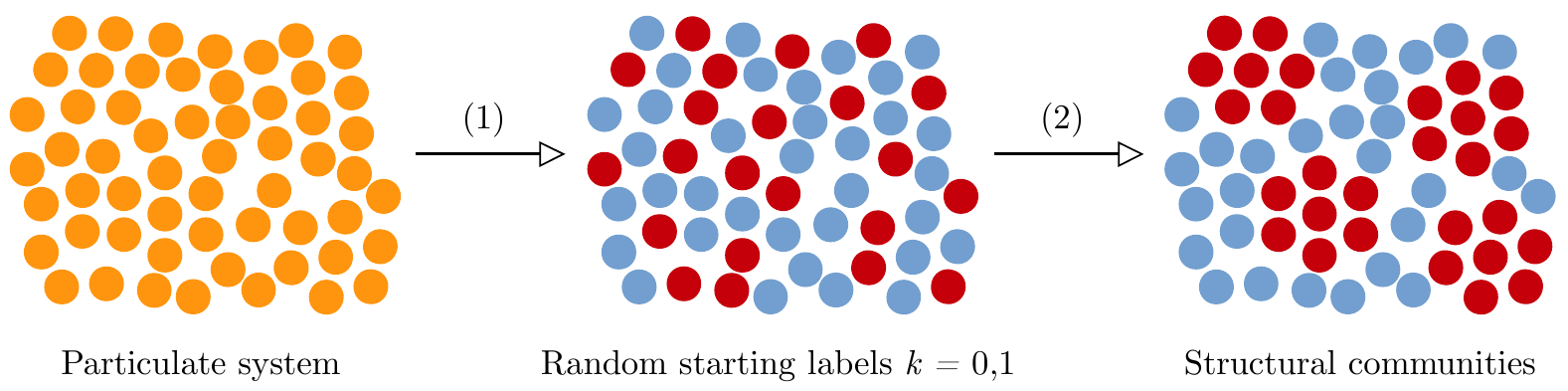}
	\caption{Schematic description of the community inference algorithm with $\mathcal{K}=2$ communities. (1) Labels $k$ are randomly assigned to each particle in the system. As a consequence, $k$ and $s$ are uncorrelated and the associated community information is close to zero, \textit{i.e.}, $I(k;s) \approx 0$. (2) Labels $k$ are stochastically reassigned following an acceptance-rejection rule that maximizes the community information. The final configuration gives the structural communities. In this sketch, the ordered community ($k=1$, red particles) corresponds to locally hexagonal packing.}
	\label{fig:procedure}
\end{figure*}

\subsubsection{Inference based on interparticle distance}
\label{sec:Ikr}

This section outlines the method by which communities are identified using the distance $r$ from the neighboring particles.
For each particle (say $i$), we identify all other neighboring particles within a distance $R$.
We compute the distances $r_{ij}$ between particle $i$ and its neighbors $j$. Let $n_i(r_m)$ be the number of neighbors $j$ of particle $i$ for which $r_{ij}$ is between $r_m= m \Delta r$ and $r_{m+1} = (m+1)\Delta r$. Here $\Delta r$ has the interpretation of a bin width in a histogram. We then define an \emph{empirical} distribution for particle $i$ as
\beq
\tilde{p}_i(r_m) = \frac{ n_i(r_m) }{ N_i } ,
\eeq
where $N_i$ is the total number of neighbors for particle $i$.
The notation with a tilde indicates that the distribution $\tilde{p}_i$ is empirical: it is computed directly from data.  This is in contrast to probability density functions such as $p(k,s)$ whose notation has no tilde: these are not computed directly, but have to be inferred.

The communities are defined such that particles in the same community have similar empirical distributions.
To this end, suppose that we have $N$ particles in total and that particle $i$ is a member of community $k_i$.
Then the empirical distribution for community $k$ is obtained by averaging over the particles in that community:
\beq
\tilde{p}_k(r_m) = \frac{1}{f_k N} \sum_{i=1}^N \tilde{p}_i(r_m) \delta_{k,k_i} ,
\label{equ:gt-sigma}
\eeq
where the Kronecker $\delta$ restricts the sum to particles $i$ in community $k$ and $f_k$ is the fraction of particles in that community.
To avoid ambiguity in notation arising from Eq.~\eqref{equ:gt-sigma}, we consistently use $k$ to indicate a community and $ij$ to indicate particles.
If $N$ is sufficiently large and bin size $\Delta r$ is small then $\tilde{p}_k(r_m)/\Delta r \approx p_k(r_m)$ is a good approximation to the community probability density function $p_k(r)$ for the distance $r$ from a particle in community $k$ to one of its neighbors (chosen at random).
Using this distribution in Eq.~\eqref{eq:MIka} with $s=r$ gives the
MI between distances and communities, which is a first example of what we refer to as \textit{community information} (CI).
It depends on which particles are assigned to which community through the $k_i$ parameters, and it depends on the data for the various interparticle distances.
To infer the communities, the CI is maximized over the $k_i$, see Sec.~\ref{sec:num-methods}.
A schematic description of this inference procedure is depicted in Fig.~\ref{fig:procedure}.
Note that in practice it may be necessary to repeat the procedure several times, or to use some annealing strategy, to determine the global maximum of the CI.

Finally, we emphasize that we are assigning \emph{particles} to communities and that each particle has many interparticle distances associated to it, which are accounted for via the empirical distribution $\tilde{p}_i$. In the context of statistical inference, assigning particles to communities in this way corresponds to a distributional clustering problem~\cite{Dhillon2003,Faiv2010}, which is not equivalent to clustering the individual values of the interparticle distances.
As noted above, this section serves to illustrate the method. In practice, our inference of communities from interparticle distances uses a slightly different CI derived from radial distribution functions, see Sec.~\ref{sec:gr-CI} below.

\subsubsection{Inference based on bond angles}

As previously mentioned, this methodology is easily generalized for other kinds of structural data.
In particular, we use it to infer communities using data for particles' bond angles, as we now explain.

For each particle $i$, we identify as before a set of neighbors, which in this case should be in the first co-ordination shell.
For every pair of neighbors $j,j'$, we identify the bond angle as the angle $\theta$ between the two vectors $\bm{r}_j-\bm{r}_i$ and $\bm{r}_{j'}-\bm{r}_i$.
We define
\beq
\Theta = - \cos\theta
\eeq
so that the probability density of $\Theta$ is flat when the neighboring particles are distributed uniformly on a sphere.
Note that this choice is specific to three-dimensional systems.
From the set of bond angles, we construct a normalized empirical distribution for particle $i$
\beq
\label{eq:qiTheta}
\tilde{q}_i(\Theta_m) = \frac{ n_i(\Theta_m)}{ N_i } ,
\eeq
where $N_i$ is the total number of bond angles that were computed for particle $i$, and $n_i(\Theta_m)$ is the number of these bond angles whose cosine is between $\Theta_m$ and $\Theta_{m} + \Delta \Theta_m$.
We have allowed here for an empirical histogram with bins of variable width; the normalization is $N_i=\sum_{m=1}^M n_i(\Theta_m)$, where $M$ is the number of bins in the histogram.
The empirical distribution for community $k$ is
\beq
\tilde{q}_k(\Theta_m) = \frac{ \sum_i n_i(\Theta_m) \delta_{k,k_i}}{ \sum_i N_i \delta_{k,k_i} } .
\eeq
As before, if $N$ is sufficiently large and bin size $\Delta \Theta_m$ is small, then
$\tilde{q}_k(\Theta_m)/\Delta\Theta_m \approx q_k(\Theta_m)$
is a good approximation to the community probability density function $q_k(\Theta)$ for the cosine of the bond angle $\theta$ between a particle in community $k$ and two of its neighbors, chosen at random. The relevant CI is then the mutual information between $k$ and $\Theta$, which is
\beq
\label{eq:IkTheta}
I( k ; \Theta ) = \sum_{k=0}^{{\cal K}-1} \int_{-1}^1  f_k q_k(\Theta) \log \left( \frac{ q_k(\Theta) }{ q(\Theta) } \right) \mathrm{d}\Theta ,
\eeq
where $q_k(\Theta)$ is the bond angle distribution for community $k$, and $q(\Theta) = \sum_k f_k q_k(\Theta)$.
The empirical MI obtained from numerical data is maximized over the community assignments to infer communities based on bond angles (or ``angular communities'').
Note that, thanks to the reparametrization invariance of the MI~\cite{coverElementsInformationTheory2006}, it is immaterial whether Eq.~\eqref{eq:IkTheta} is evaluated from the distribution of $\Theta$ or of the bond angle $\theta$ itself.
In practice, we binned the bond angles to compute the MI and explicitly checked that reparametrization invariance holds in selected cases.

\subsubsection{Bayesian interpretation}
\label{sec:bayes}

As an additional motivation for this inference method, we note that the CIs we consider can be interpreted as log-likelihoods for a Bayesian inference problem.
Hence performing inference by maximizing the CI is equivalent to maximizing the log-likelihood.
We illustrate this by the example of bond angle distributions.
As a statistical model we suppose that $\cal K$ communities exist and that each particle is identified by a community index $k_i$.
For particles in community $k$, the bond angles are assumed to be independently and identically distributed with distribution $q_k$.
This is a coarse approximation because the bond angles are correlated in practice, but it is a useful model for this illustration.
We are provided with data for the empirical bond angle distributions of each particle but the community distributions $q_k$ are unknown, as are the community labels $k_i$.
In Appendix~\ref{app:bayes}, we explain that choosing the $k_i$ to maximize $I(k;\Theta)$ can be an interpreted as choosing the most likely statistical model, given the data.

\subsubsection{Generalization to liquid mixtures}
\label{sec:liquid-mix}

As noted above, the practical models of interest in this article are supercooled liquids that are mixtures of particles of different types, which are labeled by $\alpha={\rm A, B, }\dots$
We expect that particles of different types will have different local environments.  In fact, a simple exercise is to apply our inference method to the full set of particles, and to identify communities that correspond to the two different types.  Here we are concerned with non-trivial communities, which means that we apply our algorithm separately to the particles of each type.

As a simple generalization of the algorithm to mixtures, we split type-$\alpha$ particles into two communities, and we ignore particle types when computing the empirical distribution functions.
Taking the example of bond angle distributions, the sum over $i$ in Eq.~\eqref{eq:qiTheta} is restricted to particles of type $\alpha$, but neighboring particles of all types are included when computing the bond angles of particle $i$.
The resulting community bond angle distributions are denoted by $q^\alpha_k(\Theta)$ and we define the CI for particles of type $\alpha$ as
\beq
I^\alpha( k ; \Theta ) = \sum_{k=0}^{{\cal K}-1} \int_{-1}^1 f_k^\alpha q^\alpha_k(\Theta) \log \left( \frac{ q^\alpha_k(\Theta) }{ q^\alpha(\Theta) } \right) \mathrm{d}\Theta .
\label{equ:Ialpha-bond angle}
\eeq

The result of these computations is that only particles of type $\alpha$ are assigned to communities.
Identifying communities for the other particle type is a completely separate calculation: communities for types A, B, $\dots$ are computed independently.

\subsection{Extended community inference}

In Sec.~\ref{sec:Ikr} we defined a CI based on interparticle distances.
In liquid state theory, the distribution of interparticle distances is typically studied via the liquid RDF $g(r)$.
Here we extend our community inference scheme to work with community RDFs $g_k(r)$ in place of distributions $p_k(r)$.
This has several advantages. In particular the resulting CI is sensitive to the average number of neighbors of particles in each community, as well as their distribution of distances.
It also allows a connection between the CI and the two-body excess entropy~\cite{Baranyai1989}, and to composition fluctuations.

\subsubsection{Density fluctuations}
\label{sec:gr-CI}

The RDF of community $k$ is $g_k(r)$, \textit{i.e.}, given a particle in community $k$, the function $g_k(r)$ is defined as the probability to find another particle (of either community) at a distance $r$, relative to the ideal gas case.
If there are $\mathcal{K} = 2$ communities corresponding to distinct local structures, then we expect a significant difference between $g_0(r)$ and $g_1(r)$.
To quantify this difference, we define
\beq
\Delta S_2 = \sum_{k=0}^{{\cal K}-1} \int_0^R 4\pi r^2 \rho f_k g_k(r) \log \left( \frac{ g_k(r) }{ g(r) } \right) \mathrm{d}r ,
\label{equ:dS2}
\eeq
where $\rho$ is the total number density and $g(r) = \sum_k f_k g_k(r)$ is the total radial distribution function (independent of communities).
The upper cutoff $R$ indicates the range over which the local structure is to be analyzed.
We note that the quantity $\Delta S_2$ has a similar form to Eq.~\eqref{eq:MIka}, but while $p_k(s)$ and $p(s)$ are normalized probability densities, $g_k(r)$ and $g(r)$ are not.
Thus, $\Delta S_2$ is different in essence from an MI and accounts for additional information in the structural communities.

To obtain a numerical estimate of $\Delta S_2$, we
use the same notation as Sec.~\ref{sec:Ikr} and define an \emph{empirical RDF} for particle $i$ as
\beq
\tilde{g}_i(r_m) = \frac{ n_i(r_m) }{ w(m) } ,
\label{equ:gt-i}
\eeq
where the normalization factor
\beq
w(m) = \frac{4\pi \rho \Delta r^3}{3} [(m+1)^3 - m^3]
\eeq
is the average value of $n_i(r_m)$ for an ideal gas at density $\rho$.
This is the standard normalization when deriving an RDF from the density-density correlation function.
The empirical RDF for community $k$ is obtained by averaging over the particles in that community:
\beq
\tilde{g}_k(r_m) = \frac{1}{f_k N} \sum_{i} \tilde{g}_i(r_m) \delta_{k,k_i} .
\eeq
Here again, if $N$ is sufficiently large and bin size $\Delta r$ is small then $\tilde{g}_k(r_m) \approx g_k(m\Delta r)$ is a good approximation to the community RDF appearing in Eq.~\eqref{equ:dS2}.
The quantity $\Delta S_2$ is maximized over community assignments, as described in Sec.~\ref{sec:num-methods}, to obtain extended structural communities based on distances, or ``radial communities''.

We now give an information-theoretic interpretation of $\Delta S_2$. We identify
\beq
\nbar_k = \int_0^R 4\pi r^2\rho g_k(r) \mathrm{d}r
\label{equ:nbar-sig}
\eeq
as the average number of neighbors, within the cutoff $R$, of a particle in community $k$.  We will see that one contribution to $\Delta S_2$ comes from the fact that different communities may have different values of $\nbar_k$.
By definition, the community radial distribution function $g_k(r)$ is related to the probability density $p_k(r)$ for the interparticle distance $r$, as
\beq
p_k(r) = \frac{ 4\pi r^2\rho g_k(r) }{ \nbar_k } .
\label{equ:p-sig}
\eeq
The quantities defined in Eqs.~\eqref{equ:nbar-sig} and~\eqref{equ:p-sig} have analogues for the  whole system (independent of community): they are the average number of neighbors of a particle $\nbar = \sum_k f_k \nbar_k$; also the normalized probability density for the distance to a neighbor, $P(r) =  4\pi r^2\rho g(r) / \nbar $.  Note that
\beq
P(r) = \sum_{k=0}^{{\cal K}-1} \frac{f_k \nbar_k}{\nbar} p_k(r) \; .
\label{equ:Pr}
\eeq
Then from Eq.~\eqref{equ:dS2} we have
\begin{multline}
\Delta S_2 = \overline{n} \Bigg[ \sum_{k=0}^{{\cal K}-1} \int_0^R \frac{f_k \nbar_k}{\nbar} p_k(r) \log \left( \frac{ p_k(r) }{ P(r) } \right) \mathrm{d}r
  \\ + \sum_{k=0}^{{\cal K}-1} \frac{f_k \nbar_k}{\nbar} \log \left( \frac{\nbar_k}{\nbar} \right) \Bigg] .
\label{equ:dS2-MI}
\end{multline}
The quantity within square brackets is a sum of two positive quantities.
We explain in Appendix~\ref{app:dS2-decomp} that the first term is an MI between the community $k$ and the interparticle distance, which  we denote by $I_2(k;r)$,  see Eq.~\eqref{equ:I2-kr}.  We further explain in that section that $I_2$ is an MI constructed from the joint distribution $P(k,r) = f_k \nbar_k p_k(r)/\nbar$, which corresponds physically to picking a pair of neighboring particles at random: this is similar to the MI in Eq.~\eqref{eq:MIka1} but not exactly equivalent, because Eq.~\eqref{eq:MIka1} assumes that particles (instead of pairs) are picked at random. The second term in Eq.~\eqref{equ:dS2-MI} is a relative entropy, or Kullback-Leibler (KL) divergence, that is large if the communities have different numbers of neighbors on average.
In summary, Eq.~\eqref{equ:dS2-comp} shows that the community information $\Delta S_2$ is the sum of $I_2(k;r)$, which is an MI between communities and interparticle distances, and an explicit contribution from density fluctuations. By contrast, the approach of Sec.~\ref{sec:Ikr} is not sensitive to differences in density between communities.

To apply this method to multi-component mixtures, we proceed as in Sec.~\ref{sec:liquid-mix} and split type-$\alpha$ particles into communities, ignoring particle types when computing the empirical RDFs.
When considering communities for particles of type $\alpha$, we therefore generalize Eq.~\eqref{equ:dS2} as
\beq
\Delta S_2^\alpha = \sum_{k=0}^{{\cal K}-1} \int_0^R 4\pi r^2 \rho f_k^\alpha g^\alpha_k(r) \log \left( \frac{ g^\alpha_k(r) }{ g^\alpha(r) } \right) \mathrm{d}r
\label{equ:dS2-alpha-v1}
\eeq
where $g^\alpha$ is an RDF that is centered on particles of type $\alpha$ but includes neighbors of either type, $g^\alpha_k$ is the analogous quantity but with central particles restricted to community $k$, and $f_k^\alpha$ is the fraction of type-$\alpha$ particles in community $k$
Then $g^\alpha(r) = \sum_k f_k g^\alpha_k(r)$, just as in the single-species case.

\subsubsection{Connection between $\Delta S_2$ and two-body entropy}

The extended community inference presented above maximizes $\Delta S_2$ to determine communities that are as distinct as possible. Methods that optimize other quantities might also achieve a similar result. As a motivation for this specific choice, we connect it to the two-body excess entropy defined in liquid state theory~\cite{Baranyai1989}.

From Eq.~\eqref{equ:dS2-MI} we see that $\Delta S_2$ is large in situations where specifying the community of a particle provides information about  the number of its neighbors and their distances.  It is useful to recall that the two-body excess entropy is a negative number whose magnitude is~\cite{Baranyai1989}
\beq
|S_2| = \frac12 \int_0^\infty 4\pi r^2 \rho [ g(r) \log g(r) - g(r) + 1 ] \mathrm{d}r \, .
\label{equ:classic-s2}
\eeq
This quantity measures the extent to which $g(r)$ differs from that of an ideal gas and quantifies the strength of two-body correlations in the fluid.
Stronger correlations correspond to lower entropy.
Since $S_2$ is negative in general, a larger absolute value of $S_2$ corresponds to a more ordered system.
Then
\begin{multline}
\frac{\Delta S_2}{2} + |S_2| \approx \sum_{k=0}^{{\cal K}-1}  \frac{f_k}{2} \int_0^R 4\pi r^2  \rho \Big[  g_k(r) \log g_k(r)
\\
-g_k(r) + 1 \Big] \mathrm{d}r
\label{equ:s2-wt}
\end{multline}
where the equality is now approximate because we have replaced the upper limit in Eq.~\eqref{equ:classic-s2} by $R$.
The right hand side of Eq.~\eqref{equ:s2-wt} is the weighted sum of the absolute values of the two-body excess entropies of the communities.  It is larger than $|S_2|$ because separating the particles into communities reveals additional (many-body) correlations in the system, \textit{i.e.}, the system is more ordered than one would infer from the averaged RDF $g(r)$.  This order, which is revealed by separating the system into communities, is quantified by $\Delta S_2$.

\subsubsection{Composition fluctuations}
\label{sec:mix}

We can further extend the community inference method to account for the distribution of types among neighbors of particles in community $k$.
We do this by considering an alternative CI
\beq
\Delta S_{2{\rm p}}^\alpha = \sum_{k=0}^{{\cal K}-1} \int_0^R 4\pi r^2 \sum_\beta \rho_\beta f_k^\alpha g^{\alpha\beta}_k(r) \log \left( \frac{ g^{\alpha\beta}_k(r) }{ g^{\alpha\beta}(r) } \right) \mathrm{d}r ,
\label{equ:dS2-alpha-v2}
\eeq
where $g^{\alpha\beta}$ is an RDF centered on particles of type $\alpha$, computed by considering neighbors of type $\beta$, and $\rho_\beta$ is the number density for particles of type $\beta$.
Compared to Eq.~\eqref{equ:dS2-alpha-v1}, the community information $\Delta S_{2{\rm p}}^\alpha$ now also explicitly accounts for the types of the neighboring particles.
In Eq.~\eqref{equ:dS2p-comp} of Appendix~\ref{app:dS2-decomp}, we show that $\Delta S_{2{\rm p}}^\alpha$ can be split into three pieces, analogous to the decomposition in Eq.~\eqref{equ:dS2-MI} for the single-species case. 
These are: (i) a weighted sum of conditional MIs $I_\beta^\alpha(k;r)$ that generalize $I_2(k;r)$ through a restriction to neighbors of type $\beta$; (ii) a term $I_2^\alpha(k;\beta)$ that captures the fact that different communities may have a preference for neighbors of different types; (iii) a KL divergence analogous to the second line of Eq.~\eqref{equ:dS2-MI}, which accounts for the fact that different communities may be associated with different numbers of neighbors. 
Compared with Eq.~\eqref{equ:dS2-alpha-v1}, the CI in Eq.~\eqref{equ:dS2-alpha-v2} differs through its sensitivity to the numbers of neighbors of each type [through $I_2^\alpha(k;\beta)$], and to the joint distribution of interparticle distances and types [through the $I_\beta^\alpha(k;r)$].

\subsection{Bench cases}
\label{sec:benchmarks}

\begin{figure*}[htb]
\includegraphics[scale=0.9]{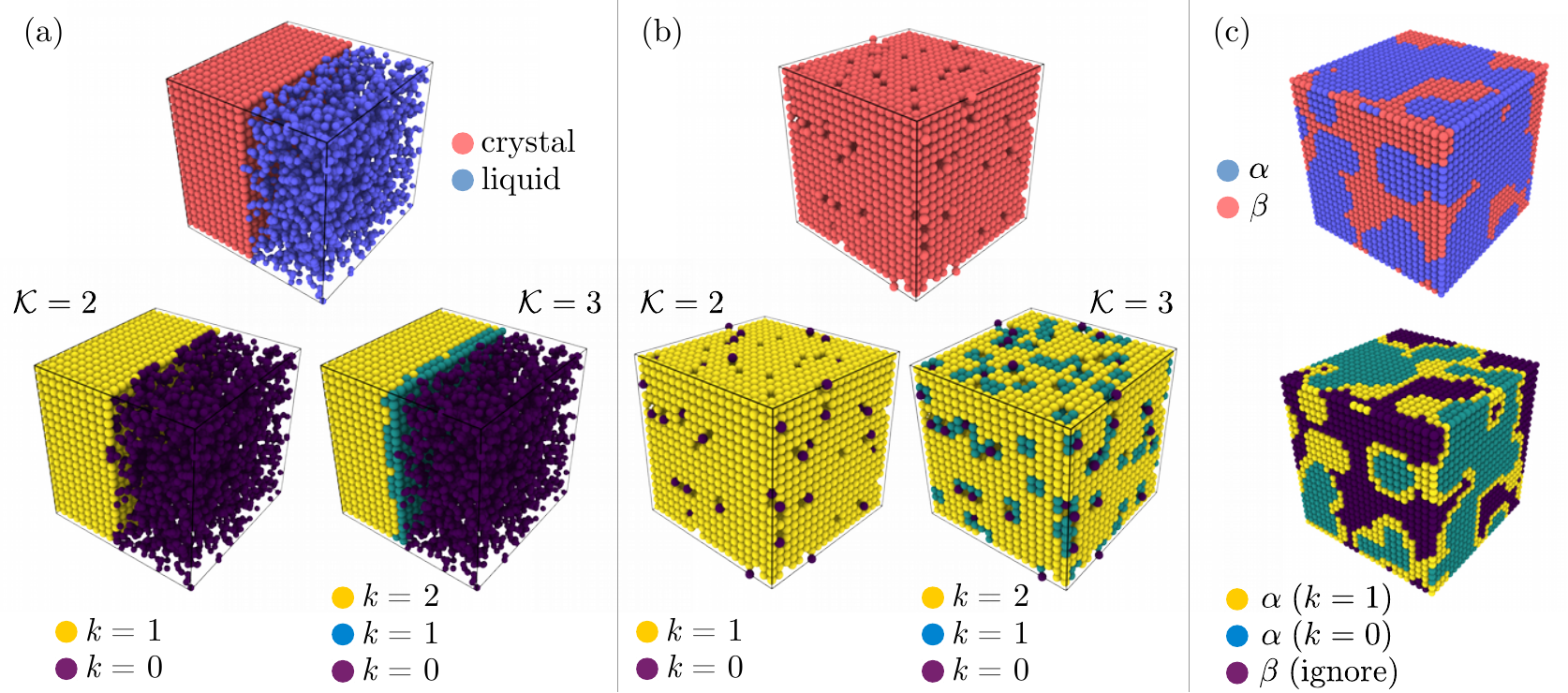}
\caption{Overview of some bench cases. (a) A perfect cubic crystal and liquid separated by an interface (top panel). Community inference with $\mathcal{K}=2$ and $\mathcal{K}=3$ gives the communities shown in the bottom left and bottom right, respectively. For $\mathcal{K}=2$, one community is identified as the crystal and the other one as the liquid. For $\mathcal{K}=3$, the additional community is identified as the interface between the two phases. (b) A perfect cubic crystal with 5\% of dislocation defects (top panel). Community inference with $\mathcal{K}=2$ (left bottom panel) identifies one community as the perfect crystal and the other one as the dislocation defects. When requesting $\mathcal{K}=3$ communities (right bottom panel), two of them are the same as the left panel and the additional one surrounds the vacancies. (c) A perfect cubic crystal of $\alpha$ and $\beta$ particles forming chemically correlated domains (top panel). The results of extended community inference using $\Delta S_{2p}^\alpha$, restricted to $\alpha$ particles and with $\mathcal{K}=2$, are shown in the bottom panel. One community is identified as the bulk of $\alpha$ particles and the other one as the interface between $\alpha$ and $\beta$ particles.}
\label{fig:benchmarks}
\end{figure*}

Before applying the method to the complex case of supercooled liquids, we conducted several tests on simpler bench cases, some of which are presented here.
When a quantitative assessment of the success of a test was not entirely straightforward, we opted for a qualitative interpretation based on visual inspection of the communities' spatial distribution.

In Fig.~\ref{fig:benchmarks}(a), we show the simple case of an interface separating a perfect cubic crystal and a dense liquid.
Using Eq.~\eqref{equ:dS2} as CI, we distinguish two separate tests: (i) for $\mathcal{K}=2$, we expect two communities coinciding with the crystal and liquid phase, respectively;
(ii) for $\mathcal{K}=3$, we expect two communities similar to (i) plus an additional community corresponding to the interface.
Test (i) is passed with high accuracy (97\% accuracy for the specific system in Fig.~\ref{fig:benchmarks}(a)). The discrepancies are obviously due to particles at the interface, where the distinction between crystal and liquid is not as well defined as for the bulk.
Concerning test (ii), simple visual inspection shows that the method successfully identifies the interface.
We note that the vast majority of the particles in the interface community belong to the crystal phase.

In Fig.~\ref{fig:benchmarks}(b) we consider the case of dislocation defects, or ``Frenkel defects'', in a perfect cubic crystal in which 5\% of the particles were randomly moved off their respective lattice sites.
Eq.~\eqref{equ:dS2} was used as CI to infer the communities. For $\mathcal{K}=2$, one of the communities is composed uniquely of defects, \textit{i.e.}, they are identified with 100\% accuracy, without false positives.
When we require $\mathcal{K}=3$, the method produces an additional community of particles surrounding the vacancies.

Finally, we designed a simple test for the extended community inference introduced in Sec.~\ref{sec:mix}, which captures composition fluctuations between communities.
To this end, we set up a perfect cubic crystal occupied by two types of particles ($\alpha$ and $\beta$) forming chemically correlated spatial domains, see Fig.~\ref{fig:benchmarks}(c).
When restricting the optimization to one particle type, say $\alpha$, we expect one community to be identified with the bulk of the $\alpha$-domains and the other one with the interface between $\alpha$- and $\beta$-domains.
In Fig.~\ref{fig:benchmarks}(c) we show that the communities obtained by optimizing $\Delta S_{2p}^\alpha$ are able to detect the interface.
By contrast, due to the lack of geometrical heterogeneity, methods that do not account for partial correlations (such as maximization of $\Delta S_2^\alpha$) do not identify relevant communities.
This test shows that an extended community inference using partial correlations can indeed prove useful in systems where the composition effects are dominant.

\section{Models and numerical methods}
\label{sec:num-methods}

We present numerical results for three binary glass-forming liquids: the Wahnstr\"om LJ mixture (Wahn)~\cite{Wahnstrom-PRA1991}, the Kob-Andersen LJ mixture (KA)~\cite{Kob-PRL1994} and a mixture of harmonic spheres (Harm)~\cite{OHern_Langer_Liu_Nagel_2002}.
These models display different kinds of locally favored structures and a varying degree of correlation between local order and dynamics~\cite{hockyCorrelationLocalOrder2014}, and are therefore well suited for our community-inference method.
We also performed several tests on simpler bench cases, some of which are discussed in Appendix~\ref{sec:benchmarks}.

The models we study consist of two species of particles, A and B, which differ in their size and their interaction parameters.
The B-particles are smaller in the Wahn and KA mixtures, whereas A-particles are smaller in the Harm mixture.
The interaction parameters of the models and their corresponding densities are given in the original papers~\cite{Wahnstrom-PRA1991,Kob-PRL1994,OHern_Langer_Liu_Nagel_2002}.
We have identified structural communities with ${\cal K}=2$ in samples composed of $N=20000$ particles. For each temperature, we have optimized simultaneously several independent configurations, as described below.

In binary mixtures, a trivial solution of the structural community inference for ${\cal K}=2$ corresponds to grouping the particles according to their type, \textit{e.g.}, $k=0$ for type A particles and $k=1$ for type B particles.
To achieve a meaningful binary partitioning into communities associated to locally ordered and disordered regions, we proceed as described in Sec.~\ref{sec:liquid-mix} and \ref{sec:mix} and carry out optimizations separately for the two species.
Another approach, which we have only partly explored in the present work, would be to request a larger number of communities ($\mathcal{K}\ge 3$) to get past the trivial partitioning by types into two separate communities.
We leave a more systematic analysis of this issue to future work.

Depending on the context, the communities are inferred by maximizing one of the information-theoretic quantities discussed in Sec.~\ref{sec:theory}.
Namely, we will consider the following two cases:
\begin{enumerate}
\item when inferring communities based on interparticle distances (``radial communities''), we use $\Delta S_2^\alpha$ from Eq.~\eqref{equ:dS2-alpha-v1} as CI;
\item for communities based on bond angles (``angular communities'') we use $I^\alpha(k;\Theta)$ from Eq.~\eqref{equ:Ialpha-bond angle} as CI.
\end{enumerate}

To identify communities, we proceed as follows.
Each particle is initially assigned a random label $k \in \{0,1\}$.
To maximize the CI, we change the label $k$ of a randomly picked particle and recompute the CI.
This new assignment is accepted if it increases the CI, otherwise it is rejected and the particle is reassigned its old label $k$.
This stochastic procedure is repeated until the system reaches a (local) maximum of the CI (see Fig.~\ref{fig:procedure}).
This gives two communities, each having its own distribution functions, \textit{i.e.}, $q_k^\alpha(\theta)$ or $g_k^\alpha(r)$.
The difference between the distribution functions quantifies the extent of structural heterogeneity in the system.
Note that $q$ is a probability density for $\Theta$, normalized as $\int_{-1}^{1}q(\Theta)\mathrm{d}\Theta = 1$; when displaying these functions we plot $q$ as a function of $\theta=\arccos{(-\Theta)}$, to facilitate identification of the preferred bond angles.

The integration cutoff $R$ in Eq.~\eqref{equ:dS2-alpha-v1} defines the length scale up to which we retain community information on the interparticle distances.
Restricting the integration to the first or second coordination shell of $g^\alpha (r)$ sometimes led to artifacts in $g_k^\alpha(r)$, such as small discontinuities due to more limited statistics.
For this reason, we decided to use a larger cutoff, up to the third coordination shell, thus including some information about medium range order in the resulting radial distribution functions.

For calculation of bond angles, nearest neighbors were identified using a fixed distance cutoff $R_{\alpha\beta}$ defined as the position of the first minimum of the relevant partial RDF $g_{\alpha\beta}(r)$. These cutoffs are $R_{AA}=1.425$, $R_{AB}=1.375$ and $R_{BB}=1.275$ for the Wahn mixture, $R_{AA}=1.425$, $R_{AB}=1.625$ and $R_{BB}=1.825$ for the Harm mixture, and $R_{AA}=1.425$, $R_{AB}=1.275$ and $R_{BB}=1.075$ for the KA mixture.

Communities are labeled at the end of the optimization according to their respective Shannon (differential) entropy
\beq
h_k[s] = - \int p_k(s) \log p_k(s) \mathrm{d}s.
\label{eq:hk-order}
\eeq
Labels are assigned using the following convention: the community with the lowest entropy is labeled as $k=1$ (more ordered) and the other one as $k=0$ (less ordered), so that $h_1[s] < h_0[s]$.
For radial communities we use Eq.~\eqref{eq:hk-order} with $p_k(s)$ replaced by $g_k(r)$. The RDF is not a normalized probability density function, so $h_k$ is not a Shannon entropy in that case. Nevertheless, it is a robust measure of structural order, related to the two-body entropy.
We found this criterion to be fairly robust in most situations.
In some cases the Shannon entropies of the structural communities have very similar values.
In such cases, the distinction between ``locally ordered'' and ``locally disordered'' communities should be taken with a grain of salt.

We perform between 50 and 100 independent optimizations (depending on temperature) for each combination of particle type (A or B) and structural measure (interparticle distances or bond angles), using different random starting labels.
The optimization with the highest CI is then kept.
In practice, most of the times our optimizations find the exact same maximum, or maxima whose values are extremely close, suggesting that the CI is a reasonably smooth function of the $k_i$.
To quantify the similarity between two optimizations $\mathcal{O}_n$ and $\mathcal{O}_m$, we define the overlap
\beq
 \mathcal{Q} = \frac{2}{N^\alpha} \sum_{i=1}^{N^\alpha} \delta_{{k_i^n},{k_i^m}} \: - 1 \; \textrm{,}
 \label{eq:overlap}
\eeq
where $N^\alpha$ is the number of type-$\alpha$ particles in the sample, $k_i^n$ is the community label of particle $i$ in optimization $\mathcal{O}_n$ and $\delta$ is the Kronecker delta.
$\mathcal{Q}=1$ corresponds to a perfect similarity between the samples (\textit{i.e.}, identical community assignments), $\mathcal{Q}=-1$ a perfect dissimilarity (\textit{i.e.}, swapped communities' labels between the two samples) and $\mathcal{Q}=0$ to community assignments that are uncorrelated between the samples.
At the lowest temperatures, the average overlap between the optimization with the highest CI and its 99 counterparts ranges from $0.93$ to $1$ depending on the model and the structural measure, in agreement with the idea of an overall convex CI landscape.

\section{Results}

\subsection{Geometry and composition of structural communities}
\label{sec:results-geom}

We start by presenting the main features of the structural communities identified by the inference algorithms detailed above.
In the following, we will consider liquids equilibrated close to the respective mode-coupling crossover temperatures~\cite{hockyCorrelationLocalOrder2014}, at which the dynamics has already slowed down by 3-4 orders of magnitude compared to the onset of slow dynamics.
Namely, the respective temperatures are $T=0.58$, $T=0.45$ and $T=5.5 \times 10^{-4}$ for the Wahnström, Kob-Andersen and harmonic spheres mixtures.
The temperature dependence of the community inference will be briefly discussed in Sec.~\ref{sec:discussion}.

\begin{figure*}[!htb]
	\includegraphics[scale=1]{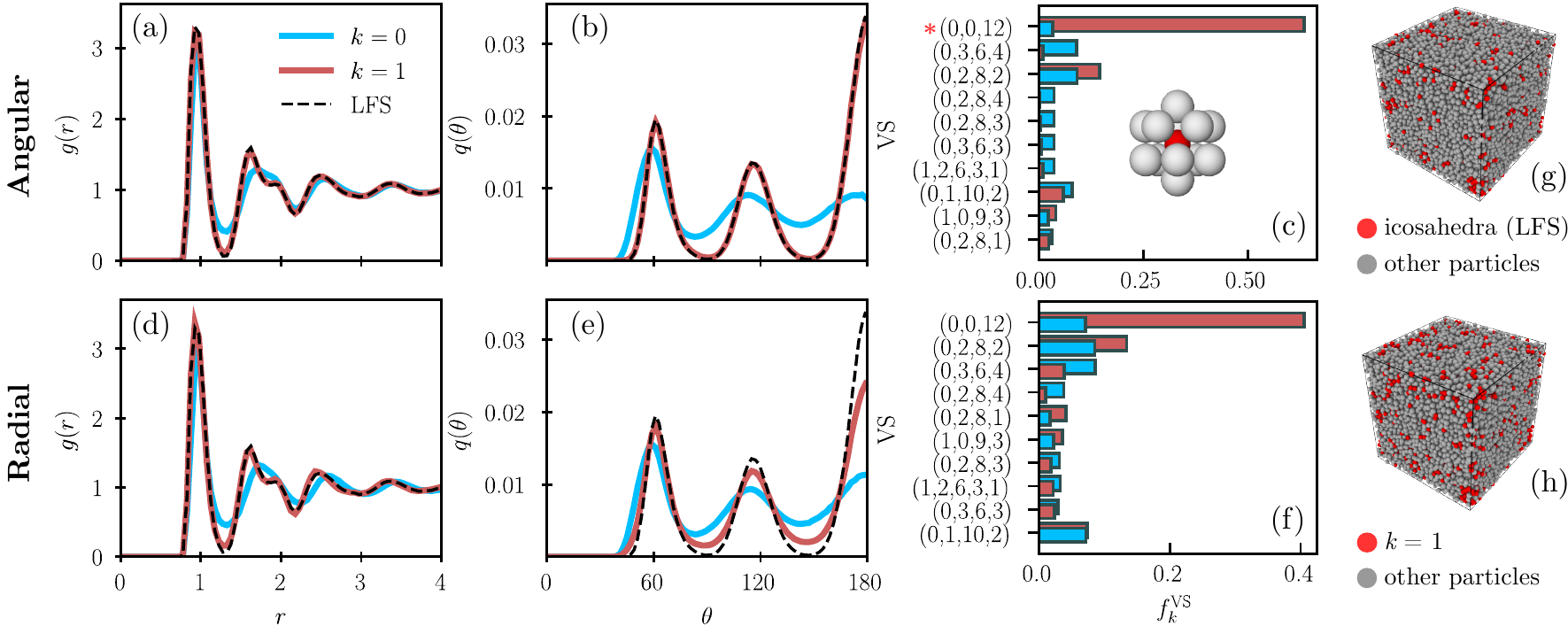}
	\caption{Features of the structural communities of type-B particles in the Wahn mixture. (a,b) Radial and angular distributions, \textit{i.e.}, $g_k^B(r)$ and $q_k^B(\theta)$, of the angular communities obtained from Eq.~\eqref{equ:Ialpha-bond angle}. The corresponding distributions restricted to particles at the center of an LFS are shown with dashed lines. (c) Fractions of the 10 most frequent VS for each angular community, by descending relative difference. The associated diversities are $D_0^B=65.7$ and $D_1^B=4.5$. The most common VS in the $k=1$ community (marked with a red asterisk) coincides with the LFS, (0,0,12), see inset. (d,e) Radial and angular distributions of the radial communities obtained from Eq.~\eqref{equ:dS2-alpha-v1}. (f) Same as (c) but for radial community. The associated diversities are $D_0^B=67.6$ and $D_1^B=13.9$. (g) Spatial distribution of the LFS in a representative sample and (h) spatial distribution of the angular $k=1$ community. All 3D visualizations were rendered in OVITO~\cite{ovito}.}
	\label{fig:wahn_all_data}
\end{figure*}

\begin{figure*}[!htb]
	\includegraphics[scale=1]{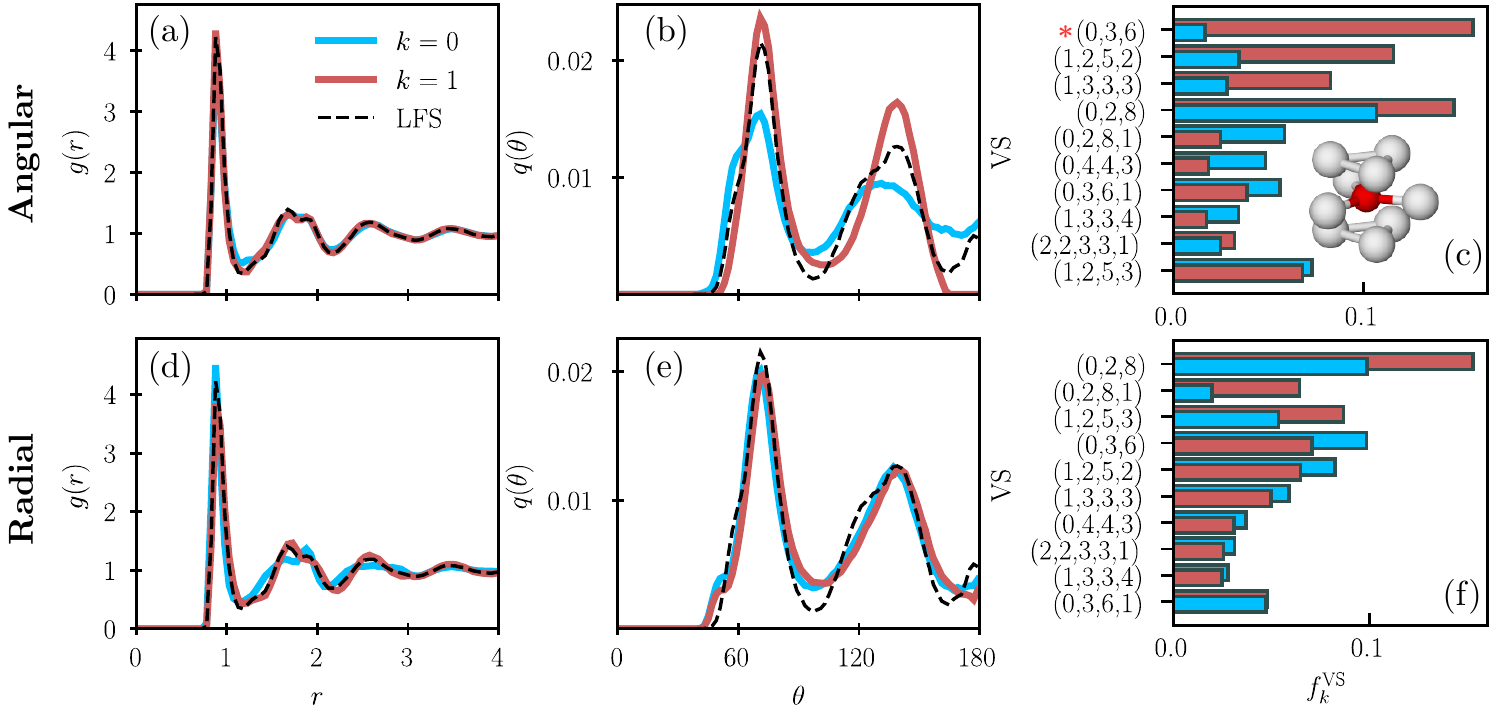}
	\caption{Features of the structural communities of type-B particles in the KA mixture. Panels (a)-(f) show the same quantities as in Fig.~\ref{fig:wahn_all_data}. Note that in this case the most common VS in the $k=1$ community, (0,3,6) (marked with a red asterisk), differs from the commonly identified LFS, (0,2,8). The (0,3,6) signature is associated to capped trigonal prismatic structures, see inset. The associated diversities are $D_0^B=61.8$ and $D_1^B=26.9$ for angular communities, and $D_0^B=52.9$ and $D_1^B=38.7$ for radial communities.}
	\label{fig:kalj_all_data}
\end{figure*}

Figures~\ref{fig:wahn_all_data}-\ref{fig:harm_all_data} provide an overview of the structural features of the communities for each given model.
In each figure, we present separately the distribution functions of the angular and radial communities, which are obtained by maximizing $I^\alpha$ and $\Delta S_2^\alpha$, respectively.
Note that only communities formed by small particles will be considered in this section, since the preferred local order in binary alloys typically develops around the smaller, usually  solute, component.
An analysis of the communities formed by the big particles can be found in Appendix~\ref{sec:big-particles}.

To provide further insight into the geometrical features of the communities, we analyze the statistics of Voronoi cells in the ordered and disordered communities.
We perform a Voronoi tessellation using the Voro++ software~\cite{voro++} and classify the local particle arrangements using the Voronoi signature (VS) of the polyhedron surrounding a given particle.
The VS of a polyhedron is defined~\cite{10.1143/PTP.58.1079} as the sequence $(n_3, n_4, \dots)$, where $n_i$ is the number of faces with $i$ vertices.
To analyze the VS composition of the communities, we compare the fractions of the 10 most common Voronoi signatures in each community and include the results in panels (c) and (f) of each figure.
We also investigate the relationship with the locally favored structures, which were identified in previous work from the statistics of the VS~\cite{coslovichLocallyPreferredStructures2011}.
In particular, the Wahn and KA mixtures are known for having preferred arrangements in the form of icosahedra and bicapped square antiprism, respectively.
The Harm mixture is characterized by distorted icosahedral structures.
The corresponding VS of these structures are (0,0,12) for the B particles of the Wahn mixture, both (0,2,8) and (1,2,5,3) for the B particles of the KA mixture, and (0,2,8,2) for the A particles of the Harm mixture. They represent respectively 22.9\%, 19.7\% and 9.7\% of the total number of small particles.
Finally, we also report a measure of the structural diversity $D_k^\alpha$ of the communities as expressed by the Shannon entropy of the distributions of the VS~\cite{weiAssessingUtilityStructure2019}.
Namely, the diversity is defined as
\beq
D_k^\alpha = \exp\left[- \sum_\mathcal{S} p(\mathcal{S}) \ln{p(\mathcal{S})}\right],
\eeq
where $p(\mathcal{S})$ is the probability of observing a Voronoi cell with VS equal to $\mathcal{S}$.
Communities $k=1$ always have a lower diversity than $k=0$, suggesting that they are indeed more (locally) ordered.
This is especially striking in the Wahn mixture, where the diversity associated to angular community $k=1$ is less than 10, close to values found in crystalline structures~\cite{weiAssessingUtilityStructure2019}.

We now briefly describe the communities in the three model systems considered.
We provide some global remarks on the nature of the structural communities in the studied models at the end of this section.

\subsubsection{Wahn model}
In the Wahn model, see Fig.~\ref{fig:wahn_all_data}(a,b), the ordered community inferred from bond angles has very similar properties to particles that form LFS.
In particular, one observes a peak in $q(\theta)$ for the ordered community at $\theta=180^\circ$.  Such a peak is a natural consequence of the inversion symmetry of the LFS.
The communities inferred from RDFs are similar, see Fig.~\ref{fig:wahn_all_data}(d,e): the bond angle distribution of particles in the ordered community differs somewhat from that of the LFS, but there is still substantial overlap.
The structural communities of the Wahn mixture are obviously connected to the pronounced icosahedral local order observed around small particles~\cite{coslovichUnderstandingFragilitySupercooled2007a}.
The snapshots in Fig.~\ref{fig:wahn_all_data}(g,h) illustrate qualitatively the striking correspondence between the ordered angular community, $k = 1$, and the particles forming icosahedral structures, \textit{i.e.}, at the center of (0,0,12) Voronoi cells.
More quantitatively, the angular community $k = 1$ includes more than 90\% of the (0,0,12) signatures.
Although angular correlations are more sensitive than radial ones in identifying local motifs, the fraction of (0,0,12) remains significant even in the radial community $k=1$ (83\%).
The ordered angular and radial communities also tolerate slight distortions of the preferred local order, as is clear from the presence of (0,2,8,2) signatures.
Structural communities thus appear robust with respect to thermal fluctuations, which can instead affect the Voronoi tessellation considerably.

\subsubsection{KA model}
In the KA model, see Fig.~\ref{fig:kalj_all_data}(a,b), the ordered community inferred from bond angle distributions also shares some features with the LFS, although it lacks the peak at $\theta=180^\circ$.
This difference may also be due to the fact that nearest neighbors identified using the fixed distance cutoff defined in Sec.~\ref{sec:num-methods} differ slightly from those determined by the Voronoi tesselation.
It is notable that the RDFs for these communities are very similar, see Fig.~\ref{fig:kalj_all_data}(a,d).
On the other hand, when the distances are used to infer communities, the resulting community RDFs differ in the second peak, while the community bond angle distributions are similar.
In fact, the communities inferred by the two methods are quite different in this model: measuring the similarity between angular and radial communities using Eq.~\eqref{eq:overlap} gives $\mathcal{Q}=-0.03$ (negligible correlation), in contrast to $\mathcal{Q}=0.41$ for the Wahn model.
The geometrical motifs of the angular communities of the KA mixture are somewhat different than expected.
We find that the ordered radial community is composed mostly by the (0,2,8) signatures, which is the bona-fide LFS of the model.
However, the ordered angular community displays different geometric features and turns out to be rich in (0,3,6) signatures, associated to capped trigonal prismatic structures.
These structures do not present linear arrangements of triplets of particles, to which optimizations based on bond angles are sensitive.
Our analysis suggests that the main ``geometric'' source of structural heterogeneity in the KA mixture comes from a different kind of motif than the bona-fide LFS.

\begin{figure*}[!tb]
	\includegraphics[scale=1]{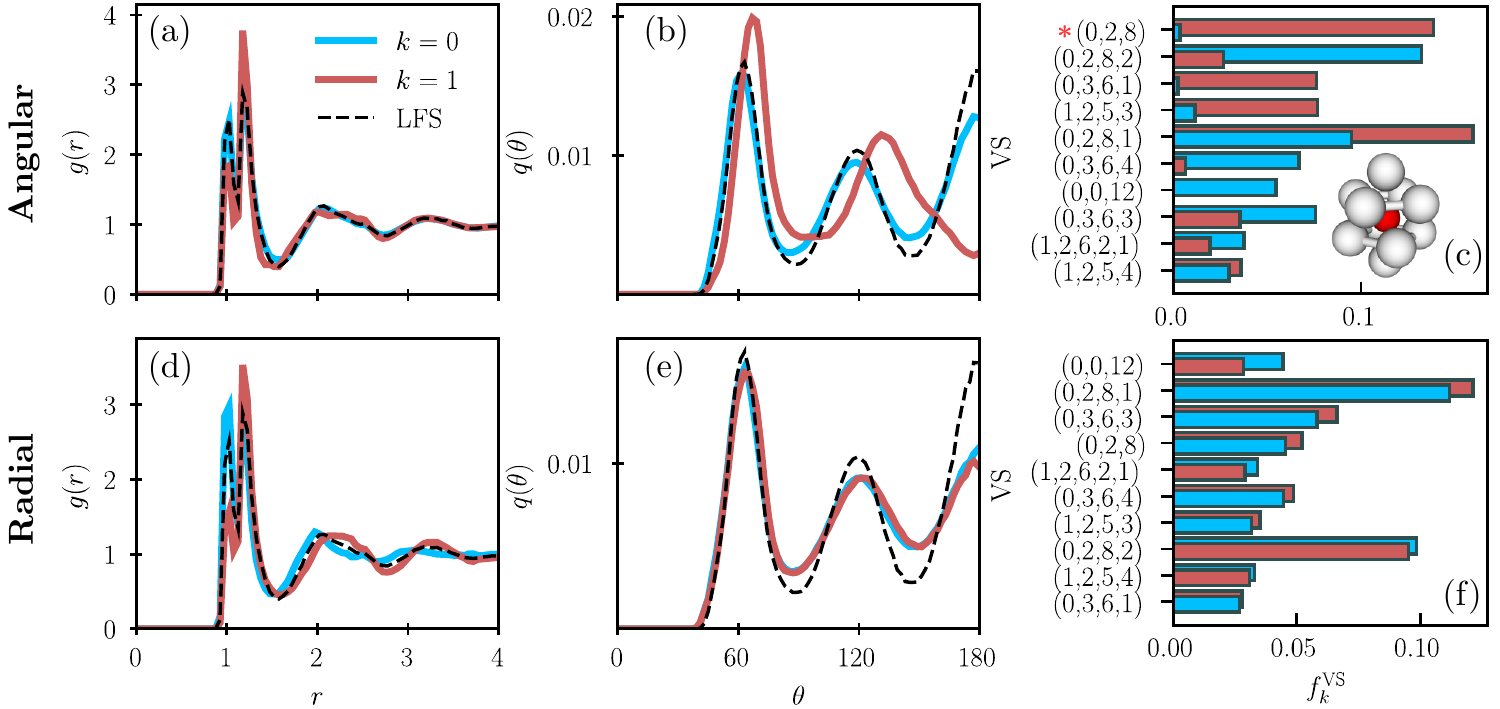}
	\caption{Features of the structural communities of type-A particles in the Harm mixture. Panels (a)-(f) show the same quantities as in Fig.~\ref{fig:wahn_all_data}. Note that in this case the most common VS in the $k=1$ community, (0,2,8,1), differs from the commonly identified LFS, (0,2,8,2). However, the largest difference in fractions between both communities is found for the (0,2,8) signature (marked with a red asterisk), which is associated to twisted square prisms, see inset. The associated diversities are $D_0^A=46.8$ and $D_1^A=33.6$ for angular communities, and $D_0^A=55.3$ and $D_1^A=52.6$ for radial communities.}
	\label{fig:harm_all_data}
\end{figure*}

\subsubsection{Harm model}
In the Harm model, one finds (perhaps surprisingly) that the disordered community ($k=0$) inferred from bond angles corresponds most closely to the LFS and exhibits a peak at $\theta=180^\circ$ as shown in Fig.~\ref{fig:harm_all_data}(a,b).
The local geometries of the two angular communities differ markedly from one another, see Fig.~\ref{fig:harm_all_data}(c).
The angular community $k = 0$ contains the vast majority of the icosahedral population (more than 90\%) and related distortions.
The icosahedral symmetry of this community is confirmed by the presence of preferred angles corresponding to typical icosahedral arrangements, see Fig.~\ref{fig:harm_all_data}(b).
By contrast, the angular community $k = 1$ contains almost the full set of (0,2,8) signatures, which we have identified again as twisted square prisms.
On average, these local structures comprise 3 particles of type A and 7 particles of type B, which slightly differs from the average coordination which is 4 for type A and 7 for type B.

It is also notable that the community RDFs for this system (centered on A particles) feature a splitting of the first peak, because the neighbors of the central particle may be of either type, see Fig.~\ref{fig:harm_all_data}(d,e).
We find that the types of particles in the first shell differ between the communities in all cases.
For communities based on bond angles, the average numbers of neighbors of each type are $\nbar_1^A = 3.9$ and  $\nbar_1^B = 7.3$ for the ordered community, and $\nbar_0^A = 5.6$ and $\nbar_0^B = 6.8$ for the disordered one.
That is, A-particles in the ordered community are preferentially surrounded by B-particles.
For communities based on RDFs we find a similar effect: $\nbar_k^A = 4.4$ and $\nbar_k^B = 7.6$ for the ordered community and $\nbar_k^A = 5.7$ and $\nbar_k^B = 6.3$ for the disordered one.  As in the KA model, the communities inferred by the two methods are very weakly correlated, their overlap is $\mathcal{Q}=0.07$.

\subsubsection{Overview and discussion}
Some general observations about this analysis are in order.
First, the presence (or absence) of a peak at $\theta=180^\circ$ in the bond angle distribution leads to a natural separation into communities.
In the models considered here, the LFS are also associated with such a peak.
These observations suggest that the number of linearly arranged triplets in a given local structure may be a simple geometric feature (along with others~\cite{tongRevealingHiddenStructural2018,Marin-Aguilar_Wensink_Foffi_Smallenburg_2019}) associated to local stability.
Second, the community distribution functions differ markedly in all models, but even more so in the Wahn mixture, for which structural heterogeneity is most pronounced.
As we shall see in Sec.~\ref{sec:discussion}, this is also reflected in the absolute values of the corresponding CI.
Finally, although the RDF of a supercooled liquid depends weakly on temperature, we find that fluctuations of the empirical RDF are significant and can be used to identify locally ordered and disordered communities of particles.
This effect is particularly pronounced in the Wahn mixture.
In the Harm and KA models, the radial communities tend to convey less geometrical information, since both the angular distribution and the distribution of VS are almost identical in the $k=0$ and $k=1$ communities.
In these two models, the difference between the communities suggests the presence of different sources of structural fluctuations, due to either density or chemical composition.

To further investigate the role of density and composition fluctuations between communities, we analyzed the communities obtained by the extended method described in Sec.~\ref{sec:mix}.
We found that maximization of $\Delta S_{2\textrm{p}}^\alpha$ produces structural communities very similar to the radial communities discussed above, obtained from Eq.~\eqref{equ:dS2-alpha-v1}.
In particular, the distribution functions of the two sets of communities were practically indistinguishable in the Wahn mixture, while some minor differences appeared in the Harm and KA models.
The overlap $\mathcal{Q}$ between the optimized communities was also fairly large, ranging from 0.5 to 0.7 depending on the model.
Thus, including explicit information on the particles' types via Eq.~\eqref{equ:dS2-alpha-v1} does not change qualitatively the nature of the communities.

However, this extended method can be exploited to disentangle more clearly the effects of composition and density.
To this end, we considered the three terms entering Eq.~\eqref{equ:dS2p-comp} and performed separate ``restricted'' optimizations by removing these terms from $\Delta S_{2\textrm{p}}^\alpha$ one at a time.
That is, instead of maximizing $\Delta S_{2\rm p}^\alpha$, we maximize other variants of the CI, such as $\Delta S_{2\rm p}^\alpha - D_\mathrm{KL}(P^\alpha(k)||f_k^\alpha)$ and $\Delta S_{2\rm p}^\alpha - I^\alpha(k;\beta)$.
We then calculated the overlaps $\mathcal{Q}$ between the optimized communities, according to which variant of the CI was used.
From this analysis, which we do not detail here, we concluded that composition fluctuations provide a larger contribution to the community information than the bare density fluctuations, because subtracting the composition term $I^\alpha(k;\beta)$ typically had a bigger effect than subtracting the density term $D_\mathrm{KL}(P^\alpha(k)||f_k^\alpha)$.
However, the nature of the communities is similar between the different variants of CI, and overlaps between them remain substantial, with $\mathcal{Q}$ ranging from 0.5 to 0.8 depending of the model.

This analysis suggests that density and composition fluctuations are strongly coupled to local structure and that the bulk of the community information is already embedded in the MIs between $k$ and $r$.
For this reason, in the following we shall not consider the extended method based on partial correlations.
However, comparing these variants of the CI may still prove useful in systems where composition fluctuations are less coupled to the local structure, \textit{e.g.}, in systems with an order/disorder transition.
A simple but explicit example is given in Sec.~\ref{sec:benchmarks}.

\subsection{Correlation between structure and dynamics}
\label{sec:corr-struct-dyn}

\begin{figure*}[!htb]
\includegraphics[scale=1]{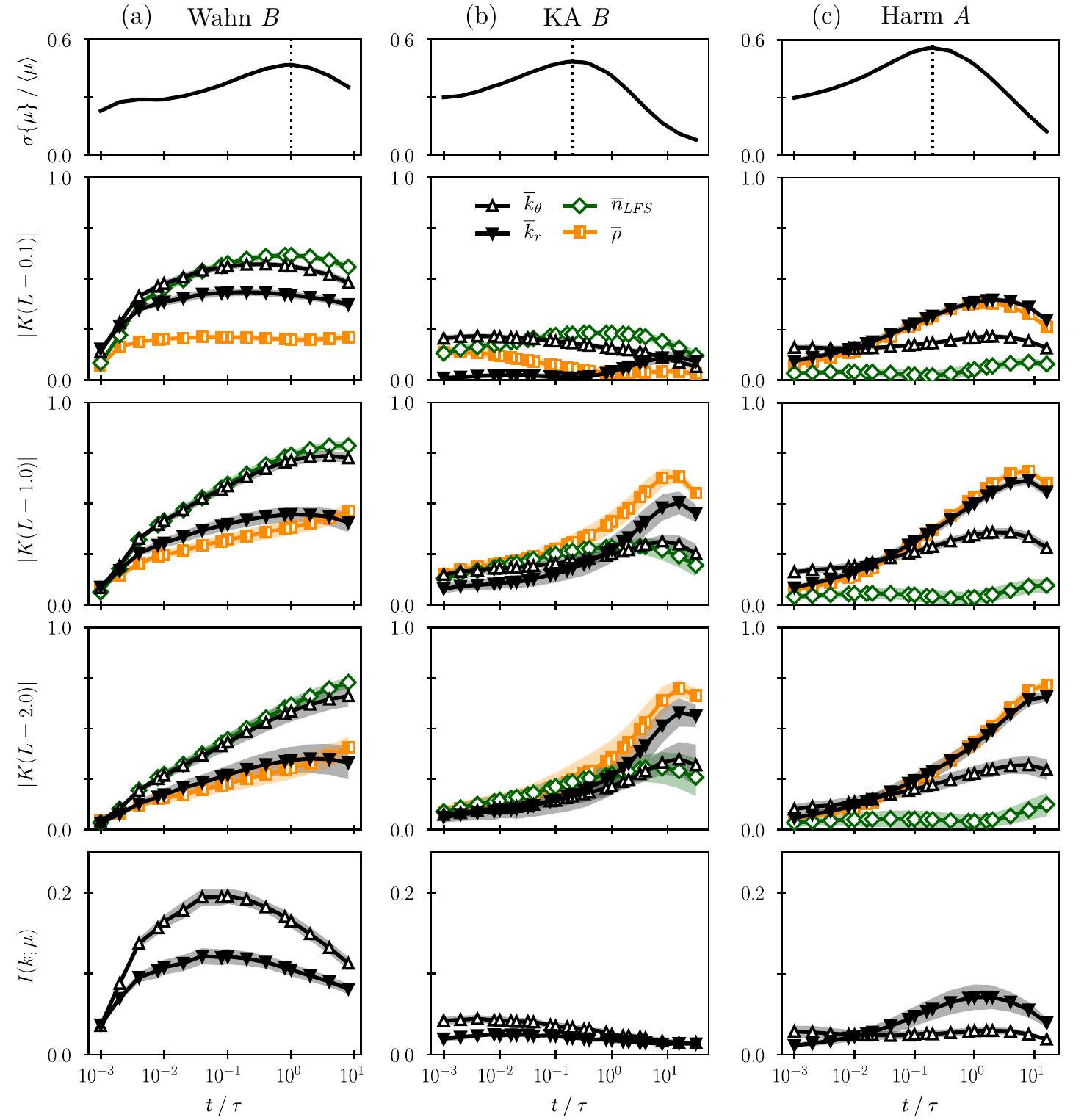}
\caption{Time-dependent correlation between the isoconfigurational mobility $\mu(t)$ and coarse-grained communities ($\overline{k}$), coarse-grained local density ($\overline{\rho}$), and coarse-grained number of LFS ($\overline{n}_\textrm{LFS}$). Columns (a), (b), and (c) show results for the small particles of the Wahn, KA, and Harm model, respectively. The top panels show the relative fluctuations $\sigma\{\mu\}/\langle \mu\rangle$ of the isoconfigurational mobility as a function of $t/\tau$. The vertical dotted lines indicate the maximum of $\sigma\{\mu\}/\langle \mu\rangle$. In the three central rows, the Spearman correlation coefficient $|K(L)|$ is shown as a function of $t/\tau$. $K(L)$ is calculated using quantities coarse-grained over the indicated length $L$. The bottom row shows the mutual information $I(k;\mu)$ between communities and isoconfigurational mobility as a function of $t/\tau$. In this figure, angular and radial communities are indicated by $\bar{k}_\theta$ and $\bar{k}_r$, respectively. Colored areas surrounding the curves correspond to the standard deviation on the distribution of the values.}
\label{fig:corr_struc_dyn}
\end{figure*}

In the previous section we found that community inference provides insight into the heterogeneity of the local structure, capturing fluctuations of geometric motifs and of composition.
Do the structural communities also correlate to the heterogeneity of the dynamics?
How predictive are they compared to other structural descriptors~\cite{hockyCorrelationLocalOrder2014}?
To address these questions, we analyze time-dependent correlations with the isoconfigurational mobility of the particles~\cite{widmer-cooperHowReproducibleAre2004}, which filters out dynamical fluctuations unrelated to structure.
For a given configuration, we perform $100$ independent molecular dynamics simulations in the microcanonical ensemble using LAMMPS~\cite{plimpton_fast_1995}, each one starting with different velocities randomly drawn from a Maxwell-Boltzmann distribution.
The isoconfigurational mobility of particle $i$ at time $t$ is then defined by $\mu_i(t) = \langle \sqrt{(\vec{r}_i(t) - \vec{r}_i(0))^2} \rangle _{IC}$, where $\langle \cdot \rangle _{IC}$ denotes the isoconfigurational average.
For each studied model, the isoconfigurational mobility was computed for 10 independent configurations.

In addition to the structural communities, we also analyze the spatial variation of the local density $\overline{\rho}$ and of the number of LFS $n_{LFS}$ around a central particle~\cite{hockyCorrelationLocalOrder2014}.
Since some of these variables are continuous and some are discrete, they will all be spatially coarse grained over a length-scale $L$.
For a given particle $i$, we define the coarse-grained local density as
$$
\overline{\rho}_i(L) = (1/L^3) \sum_{j=1}^N w(r_{ij};L) \: ,
$$
where $w(r;L)$ is a weighting function.
Similarly, for a given structural descriptor $s_i$, which can be either the number of neighboring LFS $n^{LFS}_i$ or the community label $k_i=0,1$, we define
$$
\overline{s}_i(L) = \frac{\sum_{j=1}^N s_j \cdot w(r_{ij};L)}{\sum_{j=1}^N w(r_{ij};L)} .
$$
For $\overline{s}_i(L) \equiv \overline{n}^{LFS}_i(L)$, we set $s_j=1$ if $j$ is a LFS and $s_j=0$ otherwise.
We follow Ref.~\cite{tongRevealingHiddenStructural2018} and coarse grain all structural descriptors using an exponential function $w(r;L) = e^{-r/L}$.
In the rest of this section our analysis will be restricted to correlations with the isoconfigurational mobility of the small particles, but both species are considered when coarse-graining the structural communities. Qualitatively similar trends to the ones discussed below are observed when restricting the analysis to the big particles (not shown).

Different measures of correlation between isoconfigurational mobility and structural descriptors were used in previous studies.
To establish a direct link with Refs.~\cite{hockyCorrelationLocalOrder2014,tongRevealingHiddenStructural2018}, we compute the Spearman's rank correlation coefficient $K$, which amounts to compute the Pearson correlation between the ranks of the sorted variables.
Following Ref.~\cite{jackInformationTheoreticMeasurementsCoupling2014}, we also quantify correlations using the mutual information (in bits)
\beq
I(k;\mu) = \sum_{k=0}^{\mathcal{K}-1} \int \textrm{d}\mu \: p(k, \mu) \log_2 \left( \frac{p(k,\mu)}{p(k)p(\mu)} \right) .
\eeq
Jack and co-workers~\cite{jackInformationTheoreticMeasurementsCoupling2014} suggested that in binary mixtures a strong coupling between a structural descriptor and dynamics corresponds to mutual information of the order of 0.1 bit or more.

The time-dependent Spearman correlation coefficient between $\mu_i(t)$ and the structural descriptors defined above is shown in Fig.~\ref{fig:corr_struc_dyn} for different values of the coarse-graining length $L$.
To get a feeling of how strong is the heterogeneity of the mobility field over the investigated time range, we include in the top panels the corresponding relative standard deviation of the isoconfigurational mobility, $\sigma \{\mu\} / \langle \mu \rangle$ (restricted to small particles here).
The ``contrast'' in the mobility field is strongest at the time $t^*$ at which $\sigma \{\mu\} / \langle \mu \rangle$ is maximum.
For the Wahn mixture, $t^*$ is close to the total structural relaxation time $\tau$, as obtained from the decay to $1/e$ of the total self intermediate scattering function.
For the KA and Harm mixtures, $t^* \approx 0.2 \times \tau$.

In the Wahn mixture, we find that the correlation grows in a similar way for structural communities and the LFS, and reaches large absolute values around $t^*$.
This is expected since the angular community $k=1$ and LFS are strongly overlapping.
Similar trends are observed for the radial community $k=1$.
Overall, the correlation between $\mu_i(t)$ and the structural communities of KA and Harm mixtures is weaker than in the Wahn mixture.
However, on times longer than the structural relaxation time and by increasing $L$, the correlation becomes fairly strong for both the radial communities and the local density.
In the Harm mixture, the coupling is visible even for small coarse-graining lengths $L$ for both the radial communities and the local density, but not for the angular communities.
This is consistent with the lack of locally stable geometric motifs in this model~\cite{hockyCorrelationLocalOrder2014}, at least in the accessible temperature range.

In all models, the correlation with the communities and with the local density grow with increasing length scale.
This trend suggests that the dynamic fluctuations captured by the spatially coarse-grained communities are due to a coupling with the local density.
We note that this is not a trivial result: the null hypothesis, \textit{i.e}., coarse-graining a binary random field with the same properties of the communities, leads indeed to zero correlations.
Finally, in the lower panel of Fig.~\ref{fig:corr_struc_dyn}, we show the mutual information between the isoconfigurational mobility and the structural communities.
For the B particles of the KA mixture we find that correlation is weak, which is consistent with Ref.~\cite{jackInformationTheoreticMeasurementsCoupling2014}.
The Wahn mixture shows a good correlation with an information of 0.2 bit near $t=\tau$.
The Harm mixture is in between the two other models, with an information close to 0.1 bit at $t=\tau$, suggesting that some relevant information about dynamics is indeed captured by the structural communities.

\section{Discussion}
\label{sec:discussion}

What can be learned from the application of community inference to supercooled liquids?
Perhaps the key point is that in all studied models structural communities reveal a significant heterogeneity of the local structure, hidden in the fluctuations of few-body distribution functions.
We found that strong icosahedral local order is clearly reflected in the communities of the Wahn LJ mixture.
In the other two models, structural inference provides complementary information to conventional geometrical methods, like the Voronoi tessellation.
In particular, it makes it possible to identify heterogeneity due to fluctuations in local density and composition, independent of local structure.
Explicit formulation of the inference problem in terms of partial correlations, see Eq.~\eqref{equ:dS2-alpha-v2}, provides a method to quantify the respective contribution of geometry, local density and composition to structure.
The overlap between communities inferred from Eq.~\eqref{equ:dS2-alpha-v1} and Eq.~\eqref{equ:dS2-alpha-v2} is substantial and indicates that in these models the fluctuations of geometric motifs, density and composition are indeed strongly coupled.

\begin{figure}[!tb]
	\includegraphics[scale=1]{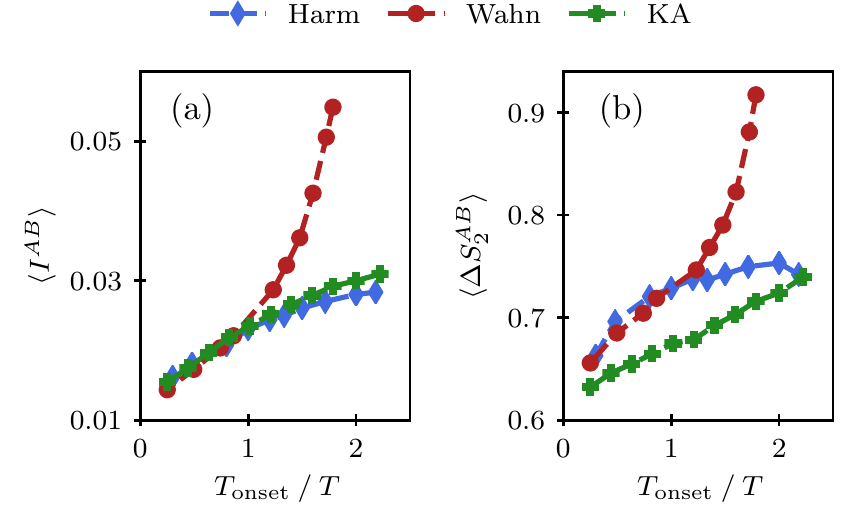}
	\caption{Temperature dependence of the ensemble average of the weighted CI for (a) angular communities and (b) radial communities. Error bars are smaller than the markers.}
	\label{fig:MI_vs_T}
\end{figure}

Our analysis has focused on temperatures close to the mode-coupling crossover temperature.
To provide some insight into how structural communities change with temperature, we analyze the variation of two simple global metrics that quantify the degree of structural heterogeneity of a liquid.
Namely, we compute the average community information, \textit{i.e.},
$I^{AB} = x_A I^A + x_B I^B$ and $\Delta S_2^{AB} = x_A \Delta S_2^A + x_B \Delta S_2^B$, where $x_A$ and $x_B$ are the chemical concentrations of the two species~\footnote{Note that the quantities $I^{AB}$ and $\Delta S_2^{AB}$ are very different from the CIs defined in Eq.~\eqref{equ:dS2} and Eq.~\eqref{eq:IkTheta}: in those cases particles of both types are treated as equivalent and the resulting CI contains no information about composition fluctuations.}.
The temperature dependence of $I^{AB}$ and $\Delta S_2^{AB}$ is shown in Fig.~\ref{fig:MI_vs_T}.
The community information increases rapidly in the Wahn mixture when the temperature drops below the onset of slow dynamics, as an expected consequence of the growth of icosahedral order~\cite{coslovichLocallyPreferredStructures2011}.
In the other two studied models, instead, the growth is mild, both for angular and radial communities \footnote{For the KA model, the position of first peak in $g_{BB}(r)$ changes appreciably with temperature. This variation has a direct consequence on the identification of the nearest neighbors, and hence on the scaling of the associated CI. We decided to use a fixed cutoff $R_{BB}=1.075$, which is suited for low temperatures, throughout the whole range of temperatures.}.
In the Harm mixture, $\Delta S_2^{AB}$ even stagnates close to the mode-coupling crossover temperature, which is reminiscent of the behavior of dynamic correlation lengths in this model~\cite{Kob_Roldan-Vargas_Berthier_2012}.
It is interesting to relate these results to the notion of fragility, as first envisaged by Angell~\cite{Angell_1991}.
Indeed, the classification of liquids into strong and fragile originally reflected ``the sensitivity of the liquids structure to temperature changes''~\cite{Angell_1995}---the coupling between structure and dynamics being assumed implicitly.
We suggest that the temperature dependence of the average community information provides a simple proxy to the concept of ``structural fragility''.
The results in Fig.~\ref{fig:MI_vs_T} are qualitatively consistent with the trends in terms of kinetic fragility, see \textit{e.g.}, Refs.~\cite{coslovichUnderstandingFragilitySupercooled2007a,Kim_Saito_2013}.

Our results also provide some insight into the long-standing problem of relating the local structure to dynamic heterogeneities.
Structural communities inferred from distances and bond angles, when coarse-grained over one or two interparticle distances, are all highly predictive of dynamic fluctuations at long times, $t>\tau$.
This is particularly true for communities based on distances, which probe the structure over an intermediate range.
Similar high correlations at long times were observed by Tong and Tanaka using a different structural order parameter~\cite{tongRevealingHiddenStructural2018}.
These observations, along with those of Ref.~\cite{jackInformationTheoreticMeasurementsCoupling2014}, indicate that it might be possible to achieve a predictive description of long-time isoconfigurational dynamics in terms of coarse-grained structural fields.
At times comparable to the structural relaxation time, instead, the correlations between communities and dynamics are system-dependent, in agreement with Ref.~\cite{hockyCorrelationLocalOrder2014}.

Community inference provides a framework to account for correlations with local density and composition fluctuations, which are sizable in the Harm mixture and which are not captured by the Voronoi tessellation.
By contrast, the KA mixture shows barely any correlation between structural communities and dynamic fluctuations on the structural relaxation time scale.
In this model, the connection between structure and dynamics is probably encoded in the energy function and eschews (at least so far) a simple geometrical interpretation.
However, it can be revealed by more complex static order parameters~\cite{coslovichStructureInactiveStates2016}.

Finally, polydisperse particles characterized by a continuous distribution of sizes constitute another class of systems where composition fluctuations are of particular interest~\cite{Coslovich_Ozawa_Berthier_2018}.
The phase behavior of hard polydisperse particles reveals a variety of complex crystalline structures, which can often be mapped to those of an effective multi-component system~\cite{Lindquist_Jadrich_Truskett_2018,Bommineni_Varela-Rosales_Klement_Engel_2019}.
It may thus be possible to describe the structural features of these systems using community inference by increasing the number of structural communities.
Preliminary results indicate that community inference with $\mathcal{K}=4$ successfully captures partial crystallization in dense, highly polydisperse hard spheres~\cite{Coslovich_Ozawa_Berthier_2018}.
It would be interesting to extend this analysis to other partially ordered systems as well as to polycrystalline materials.

\section{Conclusions}
\label{sec:conclusions}

Community inference appears as a simple and versatile tool to assess the structural heterogeneity of a physical system.
In a nutshell, the method infers $\mathcal{K}$ communities associated with a given structural property $s$ by maximizing either the mutual information $I(k;s)$ or some other measure of community information like Eq.~\eqref{equ:dS2-alpha-v1} and Eq.~\eqref{equ:dS2-alpha-v2}.
As in previous work on network-theoretic community detection~\cite{ronhovdeMultiresolutionCommunityDetection2009}, we have used a simple local optimization algorithm to find the maxima of $I(k;s)$.
It would be interesting to explore more systematically the features of the community information landscapes and the statistics of their maxima.
This may be achieved by introducing a field coupled to the community information and by exploring the community landscape through a fictitious Monte Carlo dynamics using a simulated annealing approach.
Furthermore, it would be desirable to combine different structural data, \textit{e.g.}, information on both distances and angles, and to generalize the method to a variable number of communities, which would have to be determined self-consistently as a result of the optimization.
In the physical context, future work should thus focus on these extensions as well as on the exploration of a broader set of systems.
To further develop the method, it will be also valuable to explore in more detail the relationship with existing approaches in machine learning~\cite{Dhillon2003,Faiv2010} and their potential application to the structural analysis of physical systems \cite{Reinhart2017,Boattini2019,Swanson2019,boattini2020autonomously}.

\begin{acknowledgments}
  We thank L. Berthier, W. Kob, F. Turci for useful discussions.
  We acknowledge the collaboration of A. Monemvassitis in an early stage of the work.
  Data relevant to this study will be openly available after publication at \href{https://doi.org/10.5281/zenodo.3653943}{{https://doi.org/10.5281/zenodo.3653943}~\cite{dataset}}.
\end{acknowledgments}

\begin{appendix}

\section{Bayesian interpretation of $\Delta S_2$}
\label{app:bayes}

We consider the statistical model for bond angles (or other structural data) described in Sec.~\ref{sec:bayes}.
We perform inference on data that is provided as a set of empirical distributions $\tilde{q}_i(m)$ for $i=1,2,\dots,N$ and $m=1,2,\dots,M$.
These are defined as in Eq.~\eqref{equ:gt-i}, based on the integer variables $n_i(\Theta_m)$.
Within the statistical model, each $n_i(\Theta_m)$ is an independent Poisson-distributed variable with a mean that depends on the community $k_i$ of particle $i$, and on $m$.
The corresponding mean value for $\tilde{q}_i(\Theta_m)$ is
\beq
\overline{q}_{k_i}(\Theta_m) =  \int_{m\ell}^{(m+1)\ell} q_{k_i}(\Theta) \mathrm{d}\Theta .
\label{equ:gkm}
\eeq
Then the log-probability of the data (the full set of empirical distributions) given the statistical model can be approximated as
\begin{multline}
\log P(\tilde{q}|\mathrm{model}) = \sum_i \sum_m \Bigg[ \tilde{q}_i(\Theta_m) \log \left( \frac{\overline{q}_{k_i}(\Theta_m)}{\tilde{q}_{i}(\Theta_m)} \right)
\\
-  \overline{q}_{k_i}(\Theta_m) + \tilde{q}_i(\Theta_m) \Bigg] \; .
\label{equ:P-bayes}
\end{multline}
(We used Stirling's formula to arrive at this simplified formula for the log-likelihood of these independent Poisson variables.)

The parameters of the statistical model are the distributions $p_k$, the numbers of neighbors $\overline{n}_k$, and the community labels $k_i$.  However,  $p_k$ and $\overline{n}_k$ enter only through the parameters $\overline{q}_k(\Theta_m)$.
The next step is to find the most likely values of these parameters, given the data.  This requires that we maximize $P(\mathrm{model}|\tilde{q})$ using Bayes' formula.
We take uniform priors for the community labels $k_i$ and for the parameters $\overline{q}_k(\Theta_m)$, so
$P(\mathrm{model}|\tilde{q}) \propto P(\tilde{q}|\mathrm{model})$ and it is sufficient to maximize Eq.~\eqref{equ:P-bayes} over the ${k}_i$ and the $\overline{q}$.

Extremizing first over the $\overline{q}_k$ with fixed community labels $k_i$ we find
\beq
\overline{q}_k(\Theta_m) = \tilde{q}_k(\Theta_m),
\eeq
with $ \tilde{q}_k(\Theta_m)$ defined as in Eq.~\eqref{equ:gt-sigma}.  Physically, this means that if the community labels $k_i$ are already known then we obtain the RDF for community $k$ by averaging the empirical RDFs over the particles in that community.  This is the expected result and justifies the uniform prior used for $\overline{q}$.
Hence (dropping an irrelevant additive constant that comes from normalization) we have
\begin{multline}
\log P(\mathrm{model}|\tilde{q})
= \sum_i \sum_m \Bigg[ \tilde{q}_i(\Theta_m) \log \tilde{q}_{k_i}(\Theta_m)
\\
- \tilde{q}_i(\Theta_m) \log  \tilde{q}_{i}(\Theta_m)  -  \tilde{q}_{k_i}(\Theta_m) + \tilde{q}_i(\Theta_m) \Bigg] \; .
\label{equ:Pmod-2}
\end{multline}
which is to be maximized over the $k_i$.   (Recall from Eq.~\eqref{equ:gt-sigma} that $\tilde{q}_{k_i}$ and $\tilde{q}_i$ are different physical quantities, because the subscript $k_i$ is a community index but $i$ is a particle index.)

When maximizing over the $k_i$, the term proportional to $\tilde{q}_i(\Theta_m) \log  \tilde{q}_{i}(\Theta_m)$ in Eq.~\eqref{equ:Pmod-2} is irrelevant because it does not depend on $k_i$.  In the other terms, the sums over particle indices $i$ can be simplified by partitioning particles according to their communities: one has  $\sum_i \tilde{q}_{k_i}(m) = N \sum_k f_k \tilde{q}_k(\Theta_m) = N\tilde{q}(\Theta_m)$ and $\sum_i \tilde{q}_{i}(\Theta_m)\log \tilde{q}_{k_i}(\Theta_m) = N \sum_k f_k \tilde{q}_k(\Theta_m) \log \tilde{q}_k(\Theta_m)$.  One arrives at
\beq
\frac1N \log P(\mathrm{model}|\tilde{q}) = \sum_k \sum_m f_k \tilde{q}_k(\Theta_m) \log \tilde{q}_{k}(\Theta_m)
 + C
\eeq
where $C$ is a constant that does not depend on the community labels.  Finally note that $\sum_k  f_k \tilde{q}_k(\Theta_m) \log \tilde{q}(\Theta_m)$ is also independent of the labels because $\sum_k f_k \tilde{q}_k(\Theta_m) = \tilde{q}(\Theta_m)$.  Hence
\beq
\frac1N \log P(\mathrm{model}|\tilde{q}) =\tilde{I}(k;\theta) + C'
 \label{equ:bayes-final}
\eeq
where
\beq
\tilde{I}(k;\theta) =  \sum_k \sum_m f_k \tilde{q}_k(\Theta_m) \log \left( \frac{\tilde{q}_{k}(\Theta_m) }{\tilde{q}(\Theta_m)} \right)
\eeq
and $C'$ is a new constant, independent of the $k_i$.

The quantity $\tilde{I}(k;\theta)$ is a numerical estimate of the MI $I(k;\Theta)$ defined in Eq.~\eqref{eq:IkTheta}.
In fact our numerical scheme for maximizing $I(k;\Theta)$ proceeds by maximizing $\tilde{I}(k;\theta)$.
Hence we observe that assigning community labels to maximize our numerical estimate of $I(k;\Theta)$ is equivalent to maximizing $\log P(\mathrm{model}|\tilde{g})$.
That is, the community inference method finds the most likely statistical model for the data, within the framework described in this section.

The analysis of this section may be straightforwardly generalized to other structural measures including those based on interparticle distances (as in Secs.~\ref{sec:Ikr} and \ref{sec:gr-CI}) and to liquid mixtures (as in Sec.~\ref{sec:liquid-mix}).

\section{Information theoretic analysis of $\Delta S_2$}
\label{app:dS2-decomp}

We discuss the information-theoretic content of $\Delta S_2$ and the analogous quantity $\Delta S_{2\rm p}^{\alpha}$ that takes account of partial RDFs.

\subsection{Single species}

Based on~Eqs.~\eqref{equ:nbar-sig} and \eqref{equ:p-sig} we define a joint probability distribution for $r,k$ as in the main text:
\beq
P(k,r) = \frac{f_k \nbar_k}{\nbar} p_k(r) .
\eeq
Its marginal distributions are $P(k) = \int_0^R P(k,r) \mathrm{d}r = f_k\nbar_k/\nbar$ and $P(r) = \sum_k P(k,r)$ which coincides with Eq.~\eqref{equ:Pr}.
Hence Eq.~\eqref{equ:dS2-MI} is
\beq
\frac{ \Delta S_2 }{ \nbar } = I_2(k;r) + D_\mathrm{KL}( P(k) ||  f_k ) ,
\label{equ:dS2-comp}
\eeq
where
\beq
I_2(k;r) = \sum_k \int P(k,r) \log \left ( \frac{ P(k,r) }{ P(k) P(r) } \right) \mathrm{d}r
\label{equ:I2-kr}
\eeq
is the MI associated with $P$ and
\beq
D_\mathrm{KL}( P(k) ||  f_k ) = \sum_k P(k) \log \left ( \frac{ P(k) }{ f_k } \right)
\label{equ:H-Pf}
\eeq
is a KL divergence that measures how far is the distribution $P(k)$ from its statistical null $f_k$.
This $P(k)$ is the probability that one member of a randomly chosen pair of neighbors is in community $k$.  It is larger than $f_k$ if
particles in community $k$ have more neighbors, on average.  The difference between $P(k)$ and $f_k$ also explains why the MI in Eq.~\eqref{equ:I2-kr} does not match the form of Eq.~\eqref{eq:MIka1} -- the distribution $p(k,s)$ in that definition is assumed to correspond to picking particles at random, so its marginal $p(k)$ coincides with $f_k$.

A direct generalization of the argument of this section yields a similar decomposition of Eq.~\eqref{equ:dS2-alpha-v1} in which the distribution $P$ is restricted to central particles of type $\alpha$ (and similarly $f_k$).

\begin{figure*}[!htb]
	\includegraphics[scale=1]{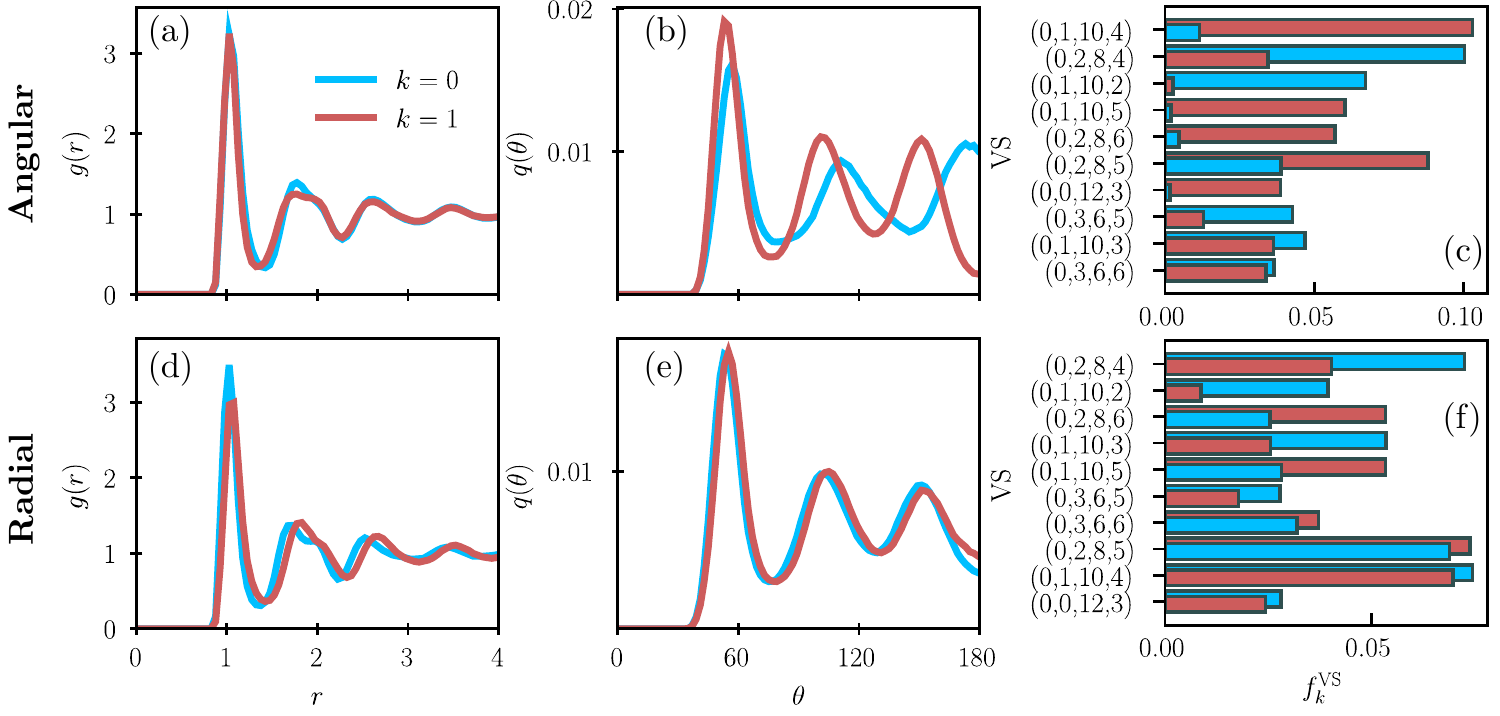}
	\caption{Features of the structural communities of type-A particles in the Wahn mixture. (a,b) Radial and angular distributions, \textit{i.e.}, $g_k^B(r)$ and $q_k^B(\theta)$, of the angular communities obtained from Eq.~\eqref{equ:Ialpha-bond angle}. (c) Fractions of the 10 most common VS for each angular community, by descending relative difference. Associated diversities are $D_0^A=83.4$ and $D_1^A=66.7$. (d,e) Radial and angular distributions of the radial communities obtained from Eq.~\eqref{equ:dS2-alpha-v1}. (f) Fractions of the 10 most common VS of each radial community. Associated diversities are $D_0^A=74.5$ and $D_1^A=94.5$.}
	\label{fig:wahn_all_data_big}
\end{figure*}

\begin{figure*}[!htb]
	\includegraphics[scale=1]{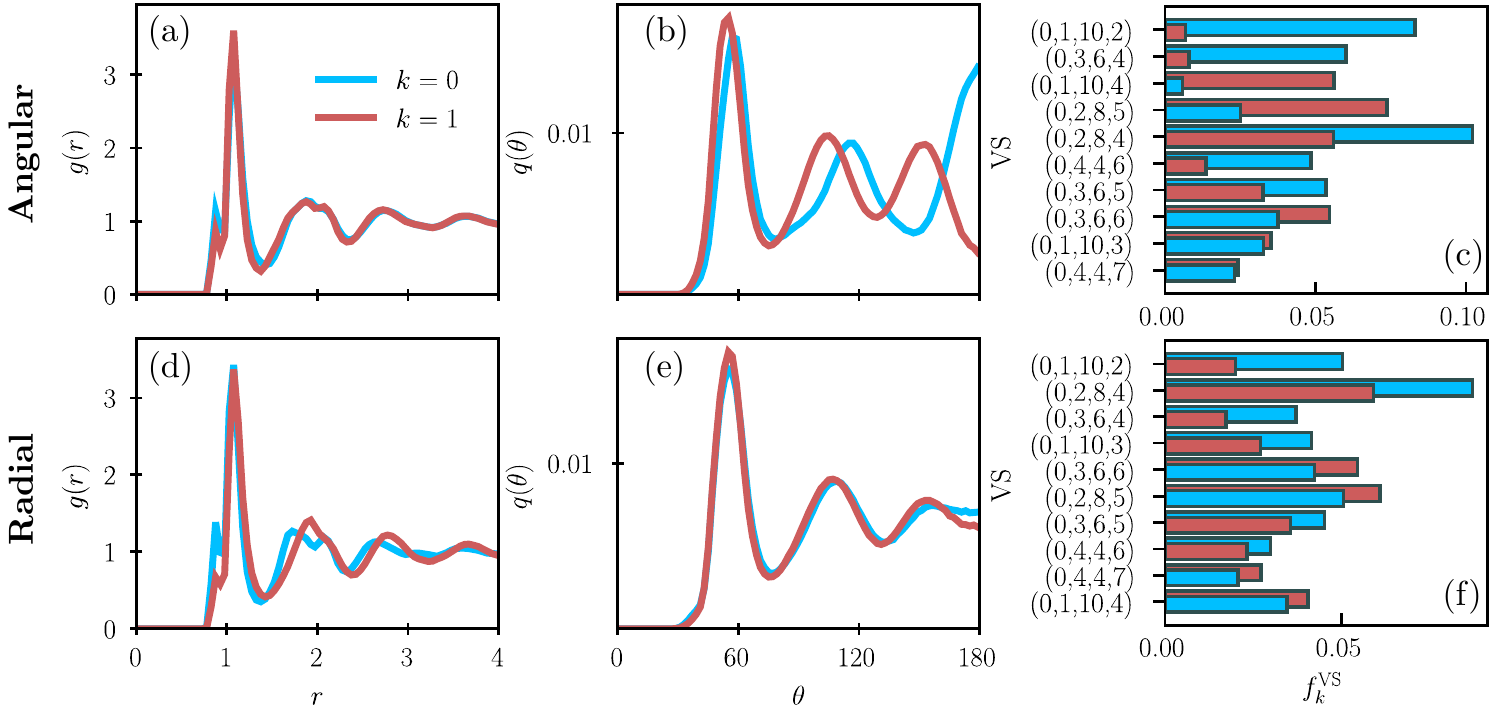}
	\caption{Features of the structural communities of type-A particles in the KA mixture. (a,b) Radial and angular distributions of the angular communities obtained from Eq.~\eqref{equ:Ialpha-bond angle}. (c) Fractions of the 10 most common VS for each angular community, by descending relative difference. Associated diversities are $D_0^A=66.6$ and $D_1^B=95.7$. (d,e) Radial and angular distributions of the radial communities obtained from Eq.~\eqref{equ:dS2-alpha-v1}. (f) Fractions of the 10 most common VS for each radial community. Associated diversities are $D_0^A=85.3$ and $D_1^A=105.2$.}
	\label{fig:kalj_all_data_big}
\end{figure*}

\begin{figure*}[!htb]
	\includegraphics[scale=1]{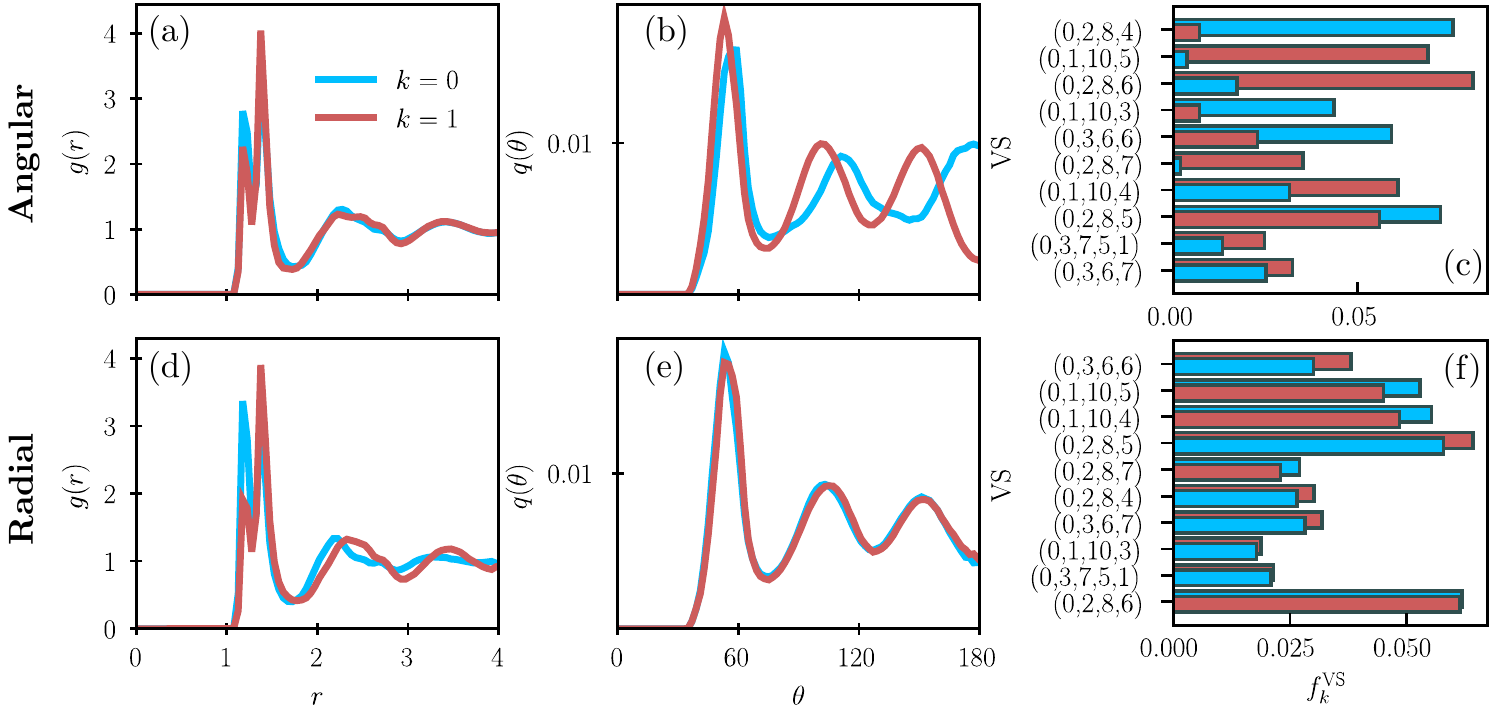}
	\caption{Features of the structural communities of type-B particles in the Harm mixture. (a,b) Radial and angular distributions of the angular communities obtained from Eq.~\eqref{equ:Ialpha-bond angle}. (c) Fractions of the 10 most common VS for each angular community, by descending relative difference. Associated diversities are $D_0^B=93.6$ and $D_1^A=91.9$. (d,e) Radial and angular distributions of the radial communities obtained from Eq.~\eqref{equ:dS2-alpha-v1}. (f) Fractions of the 10 most common VS for each radial community. Associated diversities are $D_0^B=112.3$ and $D_1^B=111.4$.}
	\label{fig:harm_all_data_big}
\end{figure*}

\subsection{Multiple species with partial RDFs}
\label{sec:splitting-DeltaS2p}

We decompose $\Delta S_{2\rm p}^\alpha$, similar to Eq.~\eqref{equ:dS2-comp}.
We work by analogy with the single-species case.  Let
\beq
p_k^{\alpha\beta}(r) = \frac{ 4\pi r^2 \rho^\beta g^{\alpha\beta}_k(r) }{ \nbar^{\alpha\beta}_k } \; ,
\eeq
with $ \nbar^{\alpha\beta}_k = \int_0^R 4\pi r^2 \rho^\beta g^{\alpha\beta}_k(r) \mathrm{d}r$ which is the average number of type-$\beta$ neighbors of a type-$\alpha$ particle in community $k$.
Analogous quantities $p^{\alpha\beta}(r)$ and $n^{\alpha\beta}$ (without community index) are defined by replacing $g_k^{\alpha\beta}\to g^{\alpha\beta}$ in these definitions.
Then Eq.~\eqref{equ:dS2-alpha-v2} becomes
\beq
\Delta S_{2{\rm p}}^\alpha = \sum_{k,\beta} \int_0^R \nbar_k^{\alpha\beta} f_k^\alpha p^{\alpha\beta}_k(r)
  \log \left( \frac{ p^{\alpha\beta}_k(r) \nbar_k^{\alpha\beta} }{ p^{\alpha\beta}(r)\nbar^{\alpha\beta} } \right) \mathrm{d}r \; .
\eeq

We now generalize the distribution $P$ from the single-species case: fix the type $\alpha$ of the central particle and define a joint distribution for $(k,r)$ and the type $\beta$ of the neighboring particle
\beq
P^\alpha(k,r,\beta) = \frac{f_k^\alpha \nbar_k^{\alpha\beta}}{\nbar^\alpha} p^{\alpha\beta}_k(r) \; ,
\eeq
with $\nbar^\alpha = \sum_\beta \nbar^{\alpha\beta}$.
We recognize $p_k^{\alpha\beta}(r) = P^{\alpha}(r|k,\beta)$ as the distribution for $r$, given the community $k$ of the central particle and the type $\beta$ of its neighbor.

As in the single-species case, the marginal for $k$ is $P^\alpha(k) = f_k^\alpha \nbar^\alpha_k / \nbar^\alpha$ and it is useful to separate in $\Delta S_{2\rm p}$
the KL divergence that accounts for differences between communities of the number of neighbors:
\begin{multline}
\frac{ \Delta S_{2{\rm p}}^\alpha }{ \nbar^\alpha } =
\sum_{k,\beta} \int_0^R P^\alpha(k,r,\beta)
  \log \left( \frac{ p^{\alpha\beta}_k(r) \nbar_k^{\alpha\beta} \nbar^\alpha }{ p^{\alpha\beta}(r)\nbar^{\alpha\beta} \nbar^\alpha_k} \right) \mathrm{d}r
\\  +
  D_\mathrm{KL}( P^\alpha(k) ||  f_k^\alpha )
  \;
  \label{equ:dS2p-int}
\end{multline}
where $D_\mathrm{KL}$ is defined as in Eq.~\eqref{equ:H-Pf}.
It can be shown that the integral in the first line of Eq.~\eqref{equ:dS2p-int} is the MI between $(\beta,r)$ and $k$, but we choose instead to decompose this term into two positive quantities (this amounts to the chain rule for MI).
We sketch the calculation.
We write:
\begin{multline}
\frac{ \Delta S_{2{\rm p}}^\alpha }{ \nbar^\alpha } =
\sum_{k,\beta} \int_0^R P^\alpha(k,r,\beta)
  \log \left( \frac{ p^{\alpha\beta}_k(r) }{ p^{\alpha\beta}(r)} \right) \mathrm{d}r
\\  +
I_2^\alpha(k;\beta) +  D_\mathrm{KL}( P^\alpha(k) ||  f_k^\alpha )
  \; ,
\end{multline}
where we used that $P^\alpha(k,\beta) = f_k \nbar^{\alpha\beta}_k/\nbar^\alpha $  as well as $P^\alpha(\beta) = \nbar^{\alpha\beta}/\nbar^\alpha$ and we identify
\beq
I_2^\alpha(k;\beta) = \sum_{k,\beta} P^\alpha(k,\beta) \log \left( \frac{ P^\alpha(k,\beta) }{ P^\alpha(k) P^\alpha(\beta) } \right)
\eeq
as the MI between the community of the central particle and the type of the neighbor.
Finally we write
\begin{multline}
\frac{ \Delta S_{2{\rm p}}^\alpha }{ \nbar^\alpha } =  \sum_\beta \frac{\nbar^{\alpha\beta}}{\nbar^\alpha} I^\alpha_\beta( k;r )
  +
I_2^\alpha(k;\beta) +  D_\mathrm{KL}( P^\alpha(k) ||  f_k^\alpha )
\label{equ:dS2p-comp}
  \; ,
\end{multline}
where
\begin{align}
I^\alpha_\beta(k;r) & = \sum_k \int P^\alpha(k,r|\beta) \log \left( \frac{ P^\alpha(k,r|\beta)} { P^\alpha(k|\beta) P^\alpha(r|\beta) } \right) \mathrm{d}r
\nonumber \\
& = \sum_k \int P^\alpha(k,r|\beta) \log \left( \frac{ p_k^{\alpha\beta}(r) }{ p^{\alpha\beta}(r) } \right) \mathrm{d}r
\label{equ:I-beta}
\end{align}
is the \emph{conditional MI} between community and distance, given that the neighbor has type $\beta$.
To get from the first to the second line of \eqref{equ:I-beta}, it is useful to note that $P^\alpha(r|\beta) = p^{\alpha\beta}(r)$.
[To recover the single-species result Eq.~\eqref{equ:dS2-comp} from Eq.~\eqref{equ:dS2p-comp}, note that $I_2^\alpha(k;\beta)=0$ in that case, and $I_\beta^\alpha$ coincides with $I_2$ in Eq.~\eqref{equ:I2-kr}.]  The physical interpretation of the three pieces in Eq.~\eqref{equ:dS2p-comp} is discussed in Sec.~\ref{sec:mix}.

\section{Community inference for the larger particles}
\label{sec:big-particles}

The analysis presented in the main text focused on the communities formed by the small particles of each model.
In Fig.~\ref{fig:wahn_all_data_big}-\ref{fig:harm_all_data_big} we present the structural features of the communities inferred for the big particles, namely type-A particles for the Wahn and KA mixtures and type-B for the Harm mixture.

As already observed for the small particles, inference of angular communities is sensitive to the presence of linear triplets of particles.
Indeed, every system shows a splitting near $\theta = 180^\circ$ between the communities' bond angle distributions $q_k^\alpha(\theta)$ (panel (b) of all three figures).
Such a marked difference is not observed in the corresponding communities' RDF $g_k^\alpha(r)$ (panels (a)), which are fairly close to one another.
Also, angular communities tend to differ significantly in terms of their local geometry, as shown by the VS composition in panels (f).
Radial communities are characterized by similar local geometries, as it is clear from the similarity of the VS distributions of the two communities.

The observations above are broadly consistent with the ones we made for communities restricted to small particles.
However, we also found that communities restricted to big particles tend to have lower absolute values of the CI (not shown) and larger values of the diversities $D_k$.
Thus, the local structure appears somewhat less heterogeneous and more disordered around the big particles than around the small ones.

\end{appendix}


%

\end{document}